\documentclass[aps,prd,reprint,superscriptaddress]{revtex4-1}
\usepackage{slashed}
\usepackage{amsmath}
\usepackage{graphicx}
\usepackage{mathtools}
\usepackage[caption=false,position=top]{subfig}
\usepackage[export]{adjustbox}
\begin{document}
%\preprint{}

\title{Fermion self-trapping in the optical geometry of Einstein--Dirac solitons}


\author{Peter E.~D.~Leith}
\author{Chris A.~Hooley}
\author{Keith Horne}
\affiliation{SUPA, School of Physics and Astronomy, University of St Andrews, North Haugh, St Andrews, Fife KY16 9SS, UK}
\author{David G.~Dritschel}
\affiliation{School of Mathematics and Statistics, University of St Andrews, North Haugh, St Andrews, Fife KY16 9SS, UK}

%\thanks{}


\date{\today}

\begin{abstract}
We analyze gravitationally localized states of multiple fermions with high angular momenta, in the formalism introduced by Finster, Smoller, and Yau [Phys Rev. D \textbf{59}, 104020 (1999)]. We show that the resulting soliton-like wave functions can be naturally interpreted in terms of a form of self-trapping, where the fermions become localized on shells the locations of which correspond to those of `bulges' in the optical geometry created by their own energy density.
\end{abstract}

%\keywords{}

\maketitle

\section{Introduction}
The interaction of quantum matter with gravity is a topic of much current interest. Since a complete picture in the form of a fully working theory of quantum gravity has yet to be formulated, analysis of specific systems is difficult, particularly in cases that exhibit strong gravitational effects. One approach, referred to as semiclassical gravity, is to approximate the full theory by keeping the gravitational field classical while treating the matter component as quantum.

We consider here systems consisting of a large number of massive, neutral fermionic particles, the mutual gravitational attraction of which results in the formation of gravitationally localized states. For fermions with finite energy, a point-like configuration would be inconsistent with the uncertainty principle, and hence these states have a non-zero extent (roughly of the order 10--100 Planck lengths), and contain no singularities. Of particular interest in this paper will be cases where the fermion mass and energy take values such that the central regions of the system become highly compressed. It is in these extreme situations that the phenomenon of fermion self-trapping becomes evident.

Here, we study such gravitationally localized states in the context of the Einstein--Dirac system, a semiclassical approximation in which the Dirac and Einstein equations are coupled. Although not a fully quantum description, in the sense that the gravitational field is treated as purely classical, and the matter content is described by a quantum wavefunction rather than a quantum field, it nonetheless can provide an interesting semiclassical description of how fermionic matter may be expected to interact with gravity. It has the advantage of solutions being more readily tractable, with the back-reaction of the matter on the space-time metric automatically included. This latter property ultimately allows for the fermion self-trapping effect to arise.

The possibility of gravitationally localized solutions of the Einstein--Dirac system was first considered by Lee and Pang in \cite{Lee1987solitonStars}, although their analysis relied on an element of approximation. It was not until 1999 that exact numerical `soliton-like' solutions were constructed by Finster, Smoller, and Yau in \cite{FSY1999original}. It is these Planck-scale, spherically symmetric, static solutions which we refer to as Einstein--Dirac solitons. These localized states have the desirable property of being free from singularities, with all metric and fermion fields being regular at the origin. In addition, the resulting space-times are asymptotically flat, and the usual Schwarzschild form is recovered outside the matter bulk, allowing a well-defined ADM mass to be extracted.

Subsequent work has been undertaken to generate analogous solutions in fermionic systems beyond the Einstein--Dirac, for example the inclusion of the electromagnetic field \cite{FSY1999maxwell}, and an SU(2) Yang--Mills field \cite{FSY2000nonAbelianBound}. Detailed analysis on black holes in this context, in particular discussion on their existence within the Einstein--Dirac system and its extensions, can be found in \cite{FSY2000bhEinsteinDirac, FSY1999bhEDM,FSY1999bhNonAbelian,Bernard2006bh}.

More recently, in \cite{Herdeiro2017bosonDiracProca} and \cite{Herdeiro2019bosonDiracProcaSpinning}, comparison has been made with the cases of boson and Proca stars, the spin-0 and spin-1 equivalents of Einstein--Dirac solitons. In this context, Einstein--Dirac solitons are referred to analogously as `Dirac stars'. The time-evolution of Dirac stars under perturbations, although at a purely classical level, has also recently been considered in \cite{Daka2019diracStarEvolution}.

Returning to the original Einstein--Dirac system, Bakucz Canário \textit{et al.} \cite{Bakucz2019powerLaw} were able to extract an analytic solution to the equations of motion, valid in the case of a massless fermion, in which all metric and fermion fields scale as simple powers of radius. Although this solution neither represents a gravitationally localized state nor is singularity-free, they were nevertheless able to demonstrate its relation to the original Einstein--Dirac solitons. In particular, the radial structure of Einstein--Dirac solitons can be understood in terms of four zones, in one of which the metric and fermion fields perform small-amplitude oscillations around this analytic `power-law' solution.

In this paper, we present gravitationally localized solutions which contain much larger numbers of particles and/or have much higher central compression than those previously studied, and in which strong gravitational effects are in evidence. We show that, in such solutions, the resulting space-time can become highly distorted, allowing a region to form containing a series of circular null geodesics (photon spheres). This can be most clearly seen by considering the optical geometry of the space-time. We go on to analyze the matter component of the solutions, showing that its behavior can be understood in terms of a fermion self-trapping effect.

The paper is organized as follows. In Sec.~\ref{secMathForm}, we describe the mathematical formulation of the problem, generalizing the original work by Finster \textit{et al.} to states with high numbers of particles, numerical results for which are presented in Sec.~\ref{secHighKappa}. In Sec.~\ref{secOptGeom}, we review the concept of optical geometry as a means of visualizing the space-time of our solutions, before describing the fermion self-trapping response in Sec.~\ref{secFermSelfTrap}. We move on to demonstrating how this self-trapping interpretation can be used to explain features in the binding energy and mass-radius plots (Sec.~\ref{secSpirals}), and to calculate the energy of the constituent fermions (Sec.~\ref{secFrequency}). In Sec.~\ref{secConclusion}, we summarize and briefly discuss the implications of our results.

\section{Einstein--Dirac system}
\label{secMathForm}

The original problem solved by Finster \textit{et al.} \cite{FSY1999original} concerned the case of two gravitationally localized fermions, the spins of which are taken to be opposite in order to satisfy spherical symmetry. To extend this analysis to states with higher numbers of fermions, while retaining the simplifications offered by spherical symmetry, it is necessary to arrange the constituent fermions in a filled shell in which the overall angular momentum is zero \cite{FSY1999bhEDM, Bakucz2019powerLaw}. Taking the total (spin + orbital) angular momentum of each individual fermion to be $j\in\{\frac{1}{2},\frac{3}{2},...\}$, the overall fermion wavefunction can be written, using the Hartree-Fock formalism, as
\begin{equation}
\Psi=\Psi_{j,k=-j}\wedge\Psi_{j,k=-j+1}\wedge...\wedge\Psi_{j,k=j},
\label{HartreeFock}
\end{equation}
where $\Psi_{jk}$ is the wavefunction of an individual fermion with angular momentum component in the $z$-direction equal to $k$. For a filled shell, the number of fermions in the state, denoted $\kappa$, is therefore equal to $2j+1$.

For large values of $\kappa$, such a single filled shell of high-angular-momentum fermions may seem somewhat less physical compared to, say, an atomic-like multiple-shell model. However, the filled shell model is sufficient to illustrate the main topic of this paper, the phenomenon of self-trapping, which is a purely gravitational effect. We might expect a similar effect to occur in the more physical multiple-shell model. 

We now provide a brief outline of the derivation of the coupled Einstein--Dirac system for such a filled shell of fermions. Throughout, we use the mostly-positive convention $(-,+,+,+)$ for the metric signature. All equations are written in natural units of $\hbar=c=1$, although factors of the Newton constant $G$ are retained. The numerical solutions presented later, however, are generated using $G=1$, allowing the radial co-ordinate to be written in units of the Planck length $l_p=\sqrt{\hbar G/c^3}$.

To derive the Dirac and Einstein equations, the starting point is the Einstein--Dirac action,
\begin{equation}
S_{ED}=\int \left( \frac{1}{8 \pi G}R + \overline{\Psi}(\slashed{D}-m)\Psi \right )\sqrt{-g}\  \mathrm{d}^4x,
\label{EDaction}
\end{equation}
the extremization of which results in the Dirac and Einstein equations:
\begin{eqnarray}
\label{DiracEqn}
\left(\slashed{D}-m\right)\Psi&=&0\,;\\ 
R_{\mu\nu}-\frac{1}{2}g_{\mu\nu}R&=&8\pi G T_{\mu\nu}.
\label{EinsteinEqn}
\end{eqnarray}
In the above, $m$ is the mass of each individual fermion, $R$ is the Ricci scalar, $R_{\mu\nu}$ the Ricci tensor, $T_{\mu\nu}$ the energy-momentum tensor, and $g=\mathrm{det}(g_{\mu\nu})$. $\slashed{D}$ is the usual Dirac operator in curved space-time, defined by $\slashed{D}=i\gamma^{\mu}\left(\partial_{\mu}+\Gamma_{\mu}\right)$, where $\Gamma_{\mu}$ is the spin connection and $\gamma^{\mu}$ are the generalizations of the Dirac gamma matrices to curved space-time, defined such that $\left\{\gamma^{\mu},\gamma^{\nu}\right\}=-2g^{\mu\nu}$.

Note that Eqs.~(\ref{DiracEqn}) and (\ref{EinsteinEqn}) form a coupled system --- the Dirac operator has an explicit dependence on the metric, and the energy-momentum tensor contains information on the matter content. As such, the Einstein--Dirac system is capable of modeling the effect of back-reaction.

Writing explicitly in the spherical co-ordinate system $(t,r,\theta,\phi)$, and following the convention introduced in \cite{FSY1999original}, we take our metric to be
\begin{equation}
g_{\mu\nu}=\textup{diag}\left(-\frac{1}{T(r)^2},\frac{1}{A(r)},r^2,r^2 \sin^2 \theta\right),
\label{metric}
\end{equation}
which is the most general form for a static, spherically symmetric system. Using this, an explicit expression for the Dirac operator can be derived:
\begin{multline}
\slashed{D}=i \gamma^t \frac{\partial}{\partial t}+i\gamma^r \left( \frac{\partial}{\partial r}+ \frac{1}{r}\left(1-\frac{1}{\sqrt{A}} \right) -\frac{T'}{2T}\right )\\+i\gamma^\theta \frac{\partial}{\partial\theta}+i\gamma^\phi \frac{\partial}{\partial\phi},
\end{multline}
where the prime represents a radial derivative.

Turning to the matter content, we require that the fermion wavefunction represent a filled shell of fermions. To this end, we take the following ansatz for each individual particle spinor wavefunction \cite{FSY1999bhEDM}:
\begin{equation}
\Psi_{jk}=e^{-i\omega t}\frac{\sqrt{T(r)}}{r}\begin{pmatrix}\chi^k_{j-\frac{1}{2}}\alpha(r)\\i\chi^k_{j+\frac{1}{2}}\beta(r) \end{pmatrix}.
\end{equation}
Note that here we are restricting our analysis to solutions with positive parity. The two-component spinor functions can be written explicitly as
\begin{eqnarray}
\chi^k_{j-\frac{1}{2}}&=&\sqrt{\frac{j+k}{2j}}Y^{k-\frac{1}{2}}_{j-\frac{1}{2}}\begin{pmatrix}1\\0\end{pmatrix}+\sqrt{\frac{j-k}{2j}}Y^{k+\frac{1}{2}}_{j-\frac{1}{2}}\binom{0}{1}\,;\\
\chi^k_{j+\frac{1}{2}}&=&\sqrt{\frac{j+1-k}{2j+2}}Y^{k-\frac{1}{2}}_{j+\frac{1}{2}}\begin{pmatrix}1\\0\end{pmatrix}-\notag \\&&\hspace{70pt} \sqrt{\frac{j+1+k}{2j+2}}Y^{k+\frac{1}{2}}_{j+\frac{1}{2}}\binom{0}{1},
\end{eqnarray}
where $Y^k_j(\theta,\phi)$ are the usual spherical harmonics. Since the solutions we seek are both static and spherically symmetric, the fermion wavefunctions are separable, with each fermion having the same energy $\omega$ and radial structure, differing only in their angular dependence. The explicit Hartree-Fock formalism is therefore not required, with the angular dependence resulting only in factors of the total particle number $\kappa$ appearing in the equations of motion. The entire matter content of the system is thus encoded in the two real fermion fields $\alpha(r)$ and $\beta(r)$.

Using the ansatz above for the metric and fermion wavefunctions, explicit expressions for the Dirac and Einstein equations can be found:
\begin{eqnarray}
\label{KappaEquations1}
\sqrt{A}\,\alpha'&=&\frac{\kappa}{2r}\alpha-(\omega T+m)\beta\,; \\
\label{KappaEquations2}
\sqrt{A}\,\beta'&=&(\omega T-m)\alpha-\frac{\kappa}{2r}\beta\,; \\
\label{KappaEquations3}
rA'&=&1-A-8\pi G \kappa\omega T^2(\alpha^2+\beta^2)\,; \\
\label{KappaEquations4}
2rA\frac{T'}{T}&=&A-1-8\pi G \kappa\omega T^2(\alpha^2+\beta^2)\notag \\[-6pt]  &&+8\pi G \frac{\kappa^2}{r}T\alpha\beta+8\pi G \kappa m T (\alpha^2-\beta^2).
\end{eqnarray}

This is a system of four coupled, 1st-order differential equations for the two metric fields $T(r)$ and $A(r)$ and the two fermion fields $\alpha(r)$ and $\beta(r)$. Equations (\ref{KappaEquations1}) and (\ref{KappaEquations2}) arise directly from the Dirac Equation, whereas Eqs.~(\ref{KappaEquations3}) and (\ref{KappaEquations4}) are the ${tt}$ and ${rr}$ components of the Einstein Equations. Note that the ${\theta\theta}$ and ${\phi\phi}$ components (equal from spherical symmetry) do not provide an additional independent equation since the Einstein equations have a vanishing covariant derivative.

\section{Einstein--Dirac solitons with large numbers of fermions}
\label{secHighKappa}

We now move on to generating localized solutions of the system (\ref{KappaEquations1})--(\ref{KappaEquations4}). We require that our solutions be asymptotically flat, i.e. both $T(r),A(r)\rightarrow0$ as $r\rightarrow\infty$. In addition, since the fermion wavefunctions are quantum mechanical, we require that solutions are correctly normalized i.e.
\begin{equation}
4\pi\int_0^\infty(\alpha^2+\beta^2)\frac{T}{\sqrt{A}}\,\mathrm{d}r=1.
\end{equation}

These conditions of asymptotic flatness and normalization are difficult to satisfy when numerically generating solutions, so we make use of the scaling procedure outlined in \cite{FSY1999original} in order to convert these into more manageable boundary conditions at $r=0$.

We also make use of the small-radius asymptotic expansion, which can be shown to take the following form for general $\kappa$ (again assuming positive parity):
\begin{align}
\label{kappaAsymptoticStart}
\alpha(r)&=\alpha_1r^{\frac{\kappa}{2}}+...\\
\beta(r)&=\frac{1}{\kappa+1}(\omega T_0-m)\alpha_1r^{\frac{\kappa}{2}+1}+...\\
T(r)&=T_0-4\pi GT_0^2\alpha_1^2\frac{1}{\kappa+1}(2\omega T_0-m)r^\kappa+... \\
A(r)&=1-8\pi G\omega T_0^2\alpha_1^2\frac{\kappa}{\kappa+1}r^\kappa+...
\label{kappaAsymptoticEnd}
\end{align}

Solutions are numerically generated using \textsc{Mathematica}'s built-in differential equation solver, NDSolve, with an explicit Runge-Kutta method. We integrate radially outwards from a small but non-zero starting radius, using Eqs.~(\ref{kappaAsymptoticStart})--(\ref{kappaAsymptoticEnd}) to calculate initial values of the fields.

Since the Einstein--Dirac system, with normalization properly applied, is inherently quantum mechanical, localized solutions occur only for a discrete number of energy values. There therefore exists a distinct ground state and a series of excited states with higher values of the fermion energy $\omega$. For the purposes of this paper, however, we restrict our analysis to the ground state, the energy of which we determine by a 1-parameter shooting procedure.

For a fixed number of particles $\kappa$, a continuous family of solutions can be found by varying the value of the central redshift $z=T(0)-1$, which gives a measure of the compression of the central regions of the soliton. Note that the value of the fermion mass is not a parameter that we can freely set; it is fixed by the choices of $\kappa$ and $z$ and determined during the shooting procedure. As such, the family of solutions generated by varying $z$ represents a set of distinct physical models in which the fermion mass differs.

For the purposes of this paper, we focus on solutions in which $\kappa$ and $z$ are both comparatively large, i.e. we consider states with a large number of particles, in which the central regions are highly compressed. Although solving the system of equations itself is no more computationally difficult, the determination of $\omega$ requires a much higher precision to be used. Our numerics therefore impose an upper limit on the value of $\kappa$ for which we can obtain solutions. The majority of results presented here are for $\kappa=90$, a value that is small enough to be computationally manageable but large enough for the self-trapping effect to be clearly evident.

\begin{figure*}
	\includegraphics[width=0.32\linewidth]{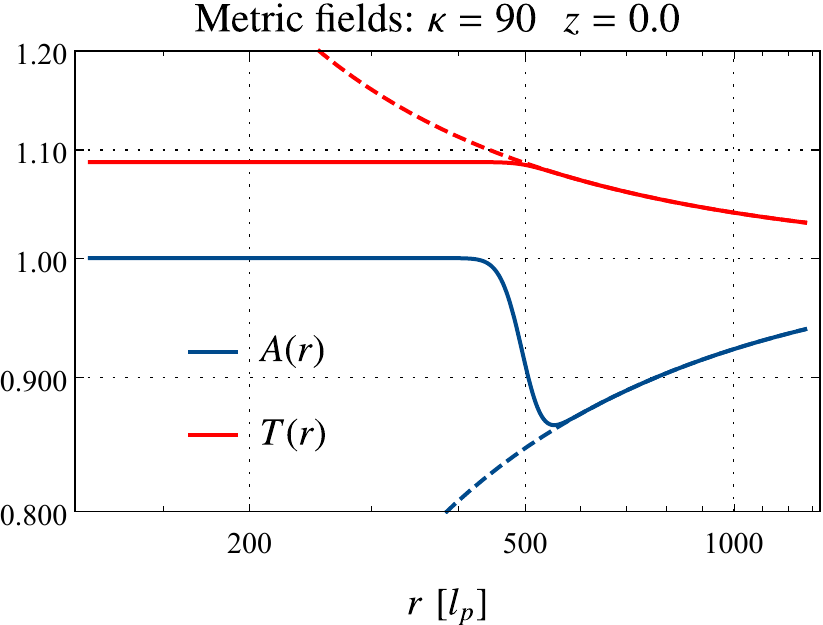}\hspace{5pt}
	\includegraphics[width=0.32\linewidth]{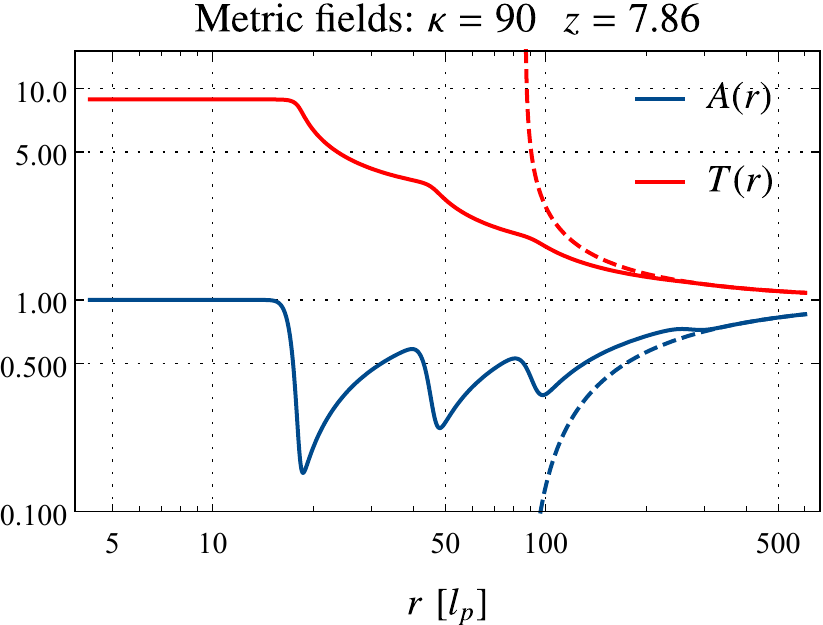}\hspace{5pt}
	\includegraphics[width=0.32\linewidth]{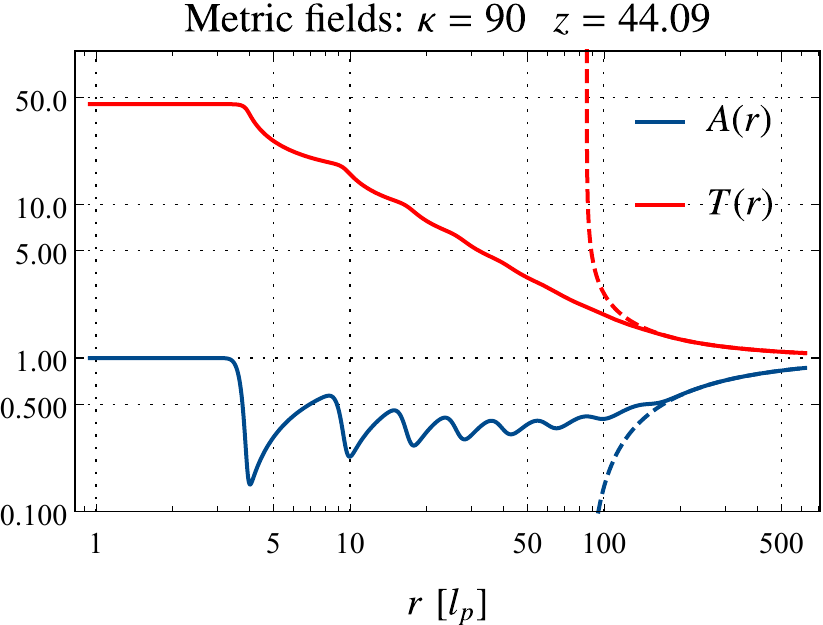}\\ 
	\vspace{5pt}
	\includegraphics[width=0.32\linewidth]{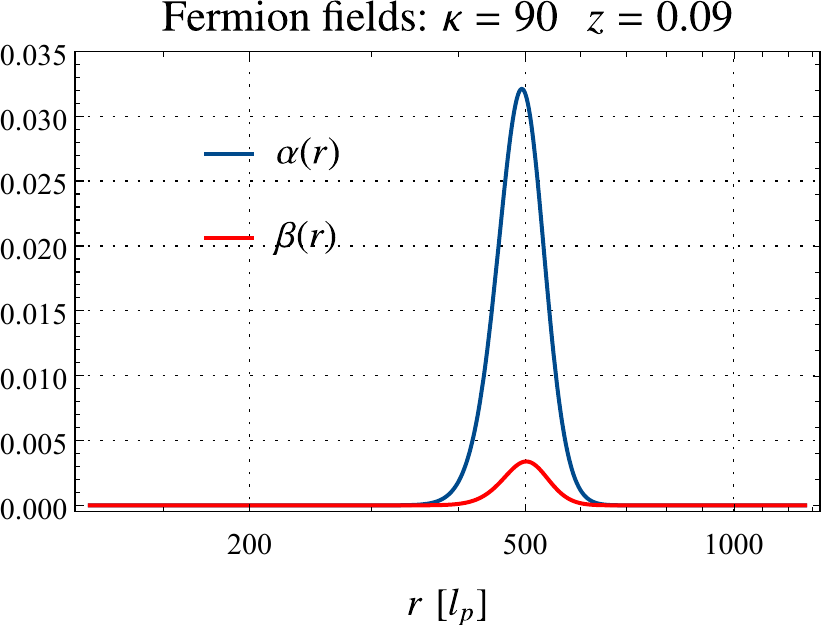}\hspace{5pt}
	\includegraphics[width=0.32\linewidth]{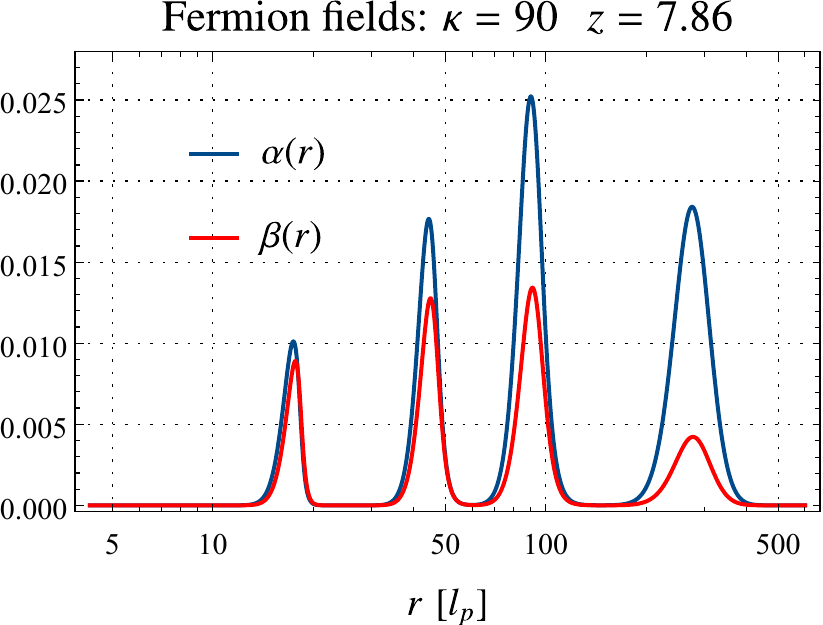}\hspace{5pt}
	\includegraphics[width=0.32\linewidth]{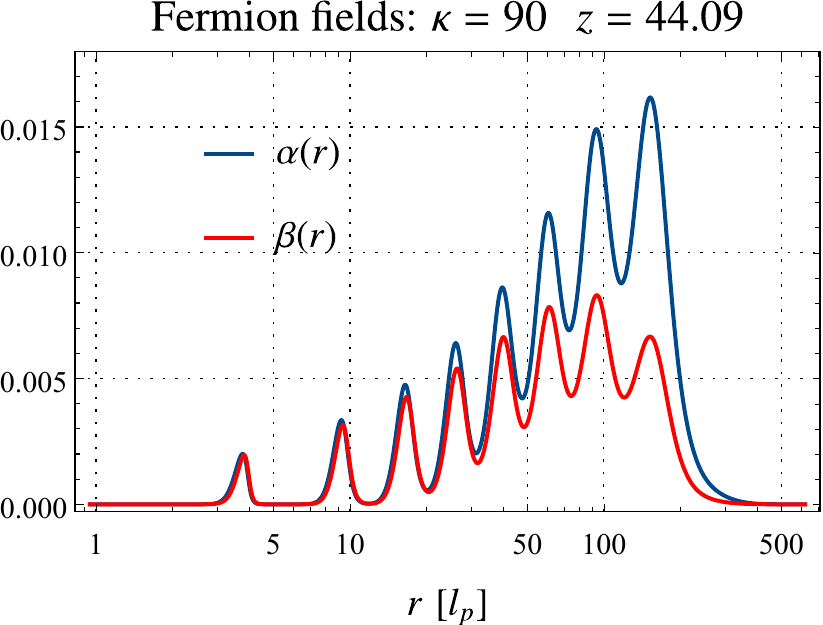}
	\caption{Plots showing the radial structure of the metric and fermion fields for solutions with $\kappa=90$ for three values of central redshift $z$. At large radii, the metric fields $A(r)$ and $T(r)$ latch on to the Schwarzschild solution with corresponding ADM mass, indicated by the dotted lines. The lowest redshift case exhibits only a single peak in $\alpha(r)$ and $\beta(r)$, with further peaks, and accompanying oscillations in $A(r)$ and $T(r)$, forming as redshift is increased. In all three cases, the value of the metric field $T(r)$ decreases monotonically with radius, a characteristic feature of Einstein--Dirac solitons.}
	\label{figHighKappa}
\end{figure*}

Solutions for three redshift cases with this value of $\kappa$ are shown in Fig.~\ref{figHighKappa}, where we plot the fermion and metric fields as a function of radius. As can be seen from these plots, the behavior of the solutions differs significantly depending on the redshift chosen. The lowest redshift case ($z=0.09$) behaves much as expected, with the fermion wavefunction exhibiting a single peak, consistent with the picture of a single filled shell of high-angular momentum fermions orbiting at a high radius. At higher redshift, however, the fermion fields split into a series of peaks, the number of which increases with $z$, with accompanying oscillations appearing in the metric fields.

It should be noted that small-amplitude oscillations in both the metric and fermion fields have been previously observed in \cite{Bakucz2019powerLaw}, for the case of $\kappa=2$, again when considering high-redshift solutions. The features seen in Fig.~\ref{figHighKappa} share similar properties with these small-amplitude oscillations, in that they appear within the `power-law' zone, and are roughly evenly-spaced in $\ln(r)$. 
We therefore suggest that they share a common origin, with the amplitude of oscillations increasing with $\kappa$, ultimately becoming large enough to result in the extreme effects shown in Fig.~\ref{figHighKappa}. In what follows, we provide a physical explanation for the appearance of these oscillations.

\section{Optical Geometry}
\label{secOptGeom}

To understand the behavior of the solutions in Fig.~\ref{figHighKappa}, we review first the concept of optical geometry. This was initially developed by Abramowicz \textit{et al.} in \cite{Abramowicz1988og}, and is most commonly utilized in the context of ultra-compact stars (see e.g. \cite{Abramowicz1997ogUltracompact} and \cite{Rosquist1999ogTrapping}).

The optical geometry approach allows for the visualization of the space-time `seen' by a null particle, by constructing a so-called optical geometry embedding diagram. This can be obtained by the following general procedure. For any spherically symmetric, static space-time with line element $\mathrm{d}s^2$, one can define a new (conformal) line element $\mathrm{d}\tilde{s}^2=(g_{tt})^{-1}\mathrm{d}s^2$, i.e. rescale the metric such that the prefactor in front of the time-component is unity. The new (conformal) time co-ordinate $\eta$ is determined by $\mathrm{d}\eta^2=(g_{tt})^{-1}\mathrm{d}t^2$. One can then perform the usual procedure of embedding this new metric in a cylindrical co-ordinate system, by rewriting the metric in the form:
\begin{equation}
\mathrm{d}\tilde{s}^2=-\mathrm{d}\eta^2+\mathrm{d}h^2+\mathrm{d}\rho^2+\rho^2\mathrm{d\varphi^2},
\label{ogSampleMetric}
\end{equation}
where $(\rho,h,\varphi)$ are the radius, height and angular co-ordinate in this new cylindrical co-ordinate system. The optical geometry embedding diagram is then defined by the surface $\rho(h)$.

\begin{figure*}
	\subfloat[Schwarzschild star]{\includegraphics[width=0.18\linewidth,valign=t]{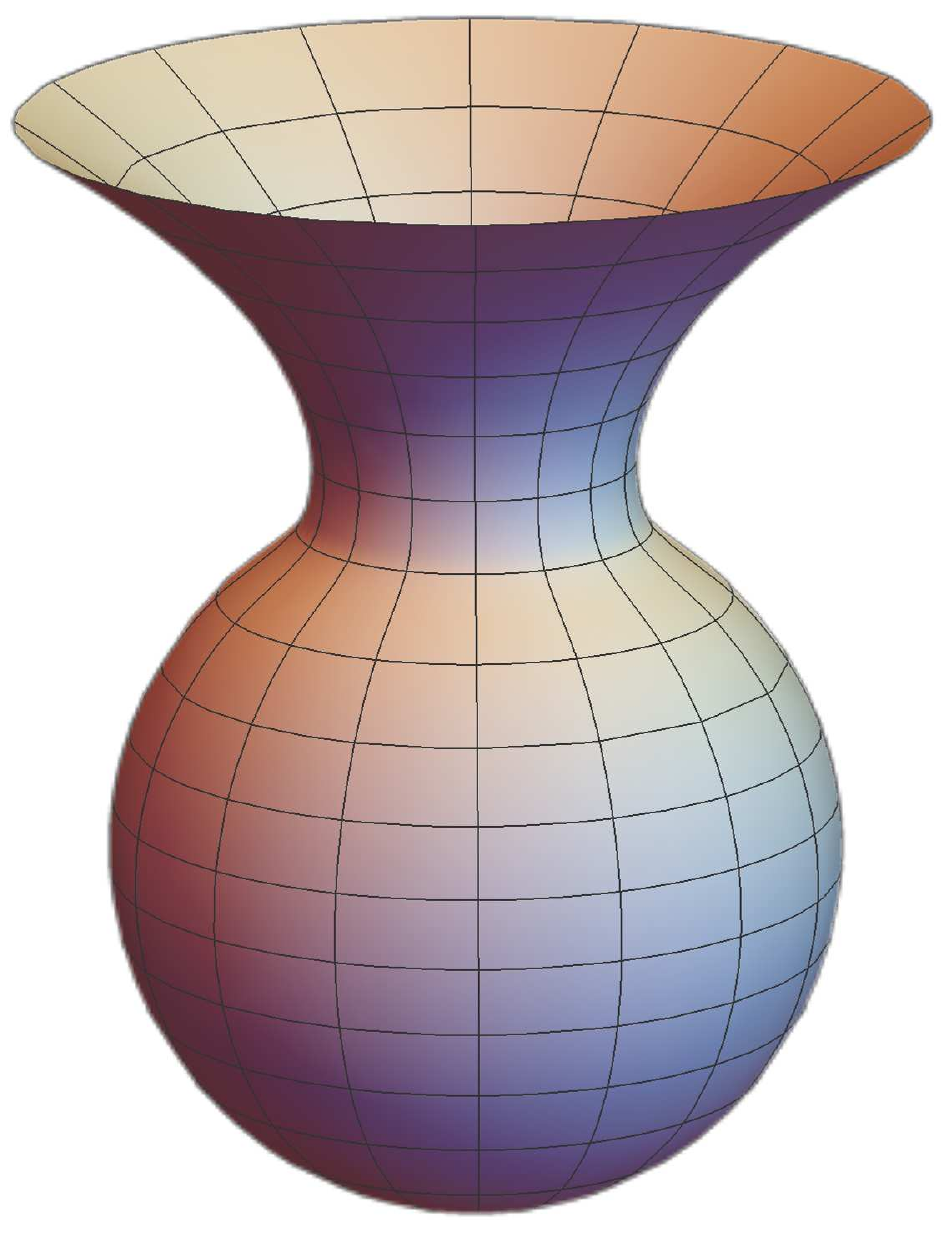}}\hspace{20pt}
	\subfloat[$z=0.09$]{\includegraphics[width=0.18\linewidth,valign=t]{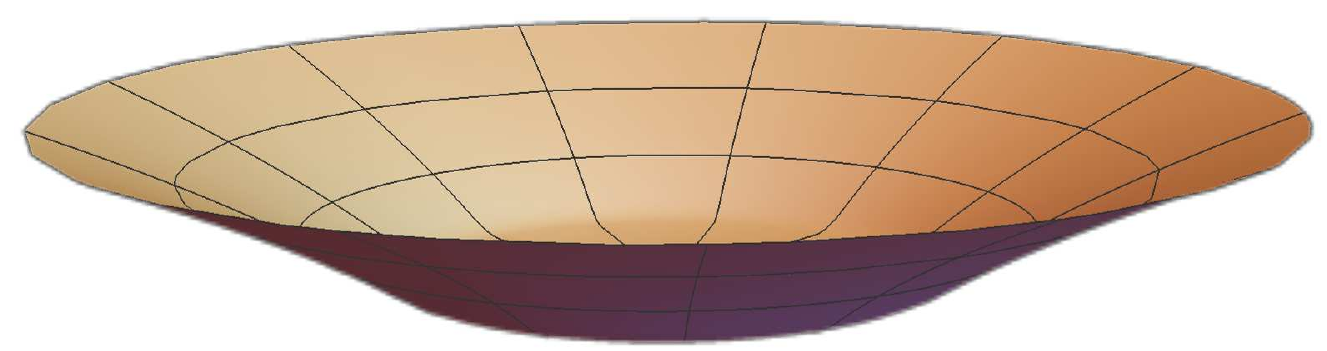}}\hspace{20pt}
	\subfloat[$z=7.86$]{\includegraphics[width=0.18\linewidth,valign=t]{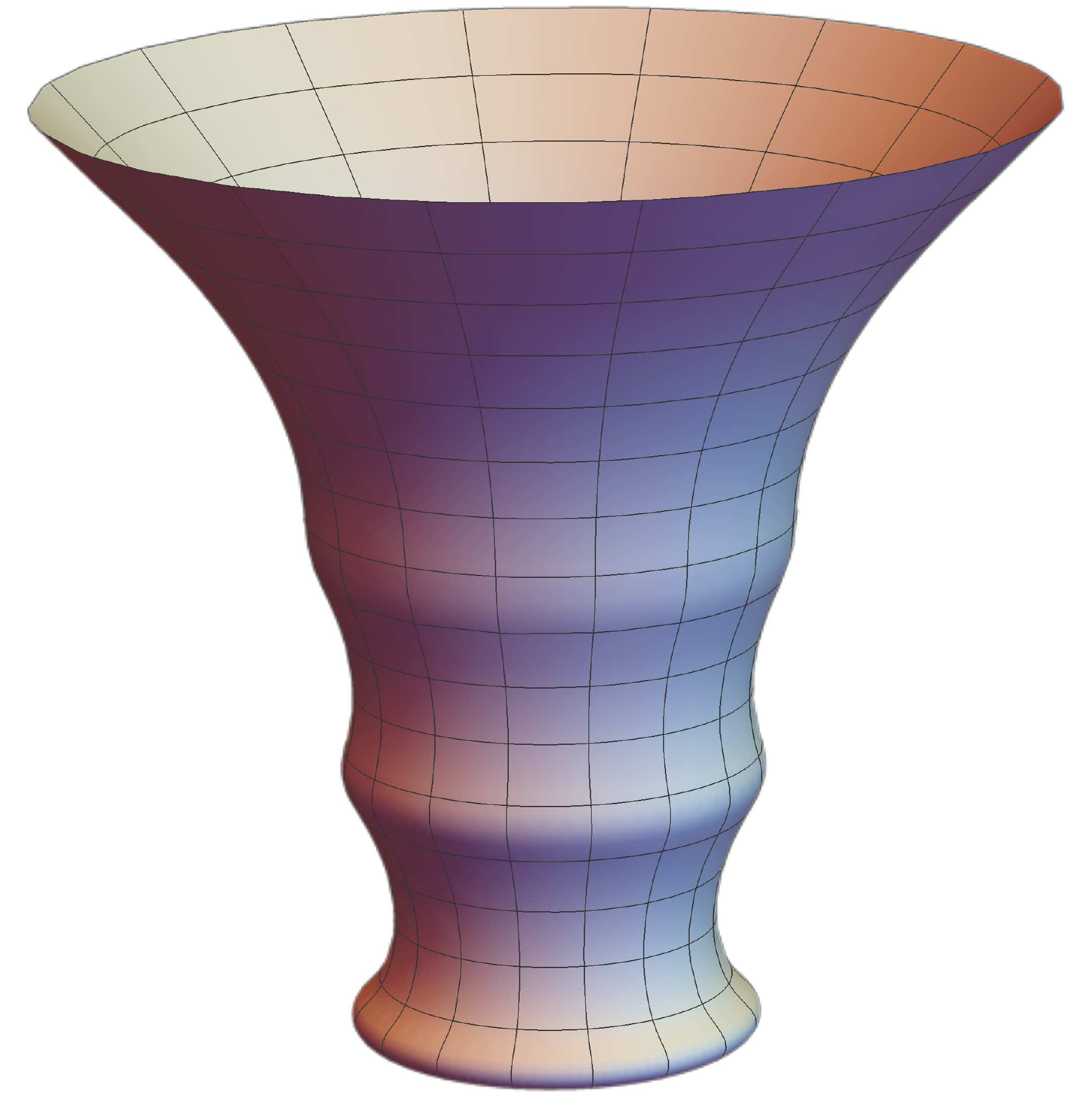}}\hspace{20pt}
	\subfloat[$z=44.09$]{\includegraphics[width=0.18\linewidth,valign=t]{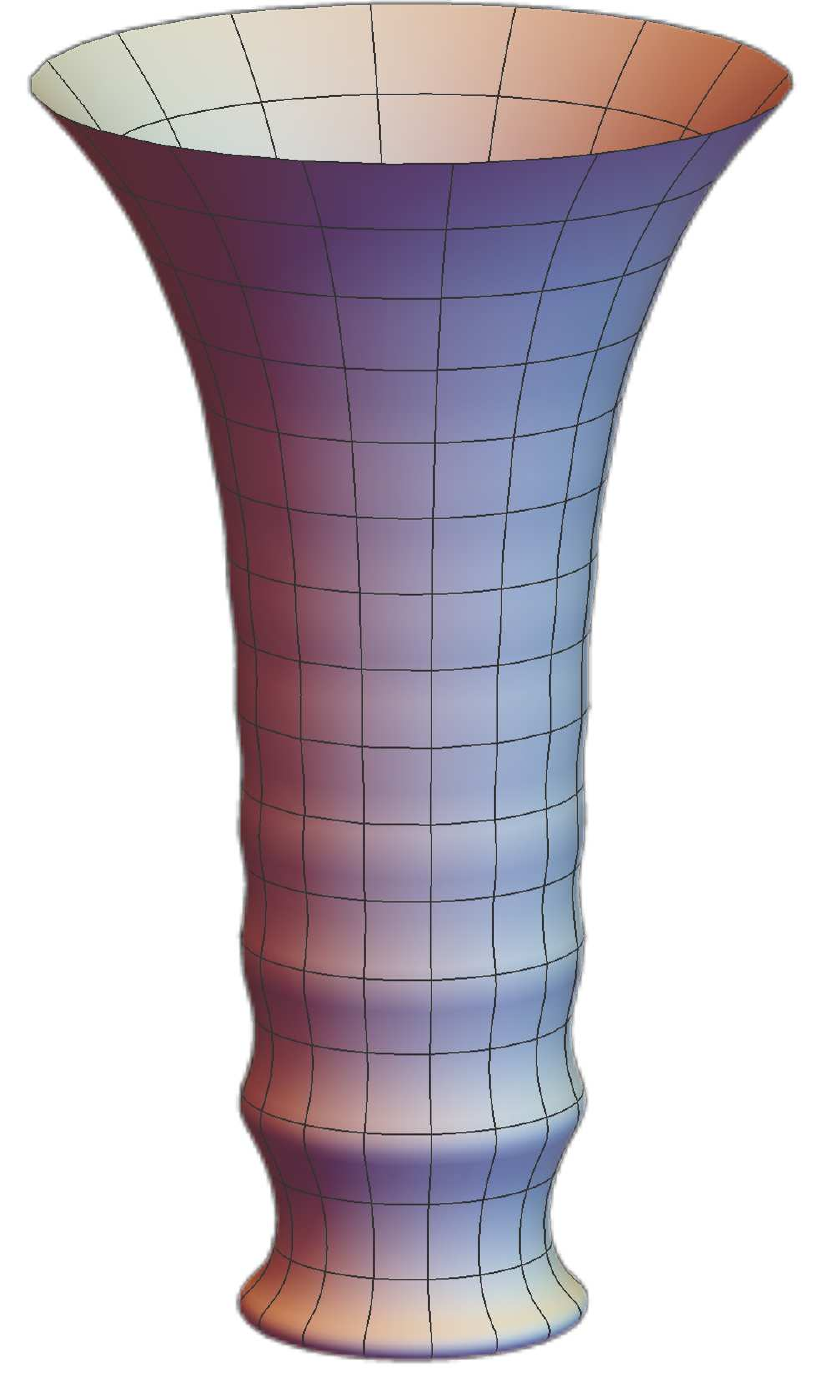}}
	\caption{Optical geometry embedding diagrams for the same three $\kappa=90$ solutions shown in Fig.~\ref{figHighKappa}, alongside that of a constant-density Schwarzschild star with $R<3M$ for comparison. The base of each diagram corresponds to $r=0$. with soliton radius $r$ increasing from bottom to top. The bulges/necks in the optical geometry correspond to stable/unstable circular null geodesics. As redshift is increased, a cylindrical structure appears in the optical geometry along which bottlenecks form, with both the length of this region and the number of bottlenecks increasing. Note that the soliton solutions have a much flatter base than the Schwarzschild star, and the bottlenecks which appear are not so pronounced. }
	\label{figOgEmbedding}
\end{figure*}

Figure \ref{figOgEmbedding}(a) reproduces the results in \cite{Abramowicz1997ogUltracompact}, showing the optical geometry embedding diagram for an ultra-compact constant density Schwarzschild star. This has the characteristic `bottle-neck' shape, where the `bulge' of the bottle-neck, located within the star, is a stable circular null geodesic (photon sphere), while the `neck' corresponds to an unstable circular null geodesic. It is argued in \cite{Abramowicz1997ogUltracompact} and \cite{Abramowicz1993bh} that the appearance of such a bottle-neck structure endows the space-time with the ability to trap null particles, owing to the reversal of the centrifugal force around the circular null geodesics. Such structures can only appear, however, if the star is sufficiently compact, causing high distortion of the resulting space-time.

To generate similar optical geometry embedding diagrams for Einstein-Dirac solitons, we first define a new metric:
\begin{equation}
\mathrm{d}\tilde{s}^2=-\mathrm{d}\eta^2+\frac{T^2}{A}\mathrm{d}r^2+r^2T^2\mathrm{d}\Omega^2.
\label{ogSampleMetric2}
\end{equation}
where $\mathrm{d}\Omega^2=\mathrm{d}\theta^2+\sin^2\theta\,\mathrm{d}\phi^2$. Then, by projecting the two redundant angular co-ordinates $\theta$ and $\phi$ onto a single cylindrical angular co-ordinate $\varphi$, and comparing Eqs.~(\ref{ogSampleMetric}) and (\ref{ogSampleMetric2}), we can identify the new radial co-ordinate as $\rho=rT$. Solving the resulting condition $\mathrm{d}z^2+\mathrm{d}\rho^2=T^2 A^{-1} \mathrm{d}r^2$ allows us to write our metric in the form of Eq.~(\ref{ogSampleMetric}), with the height of the embedded surface given by the expression:
\begin{equation}
h(r)=\int_0^{r}{T(u)\sqrt{\frac{1}{A(u)}-1-\frac{rT'(u)}{T(u)}}\ \mathrm{d}u}.
\end{equation}

Plots of the optical embedding diagrams for three of our solutions, with $\kappa=90$ and differing redshift, are shown in Figs. \ref{figOgEmbedding}(b)-(d). These correspond to the three solutions shown in Fig.~\ref{figHighKappa}. For the lowest redshift case ($z=0.09$), the optical geometry is of a simple `saucer' shape, opening out from a base point into an exterior Schwarzschild metric. As redshift is increased, however, we see the appearance of a tubular structure, along which a series of necks and bulges, each of which corresponds to the location of a circular null geodesic. The number of these bottleneck structures is directly related to the number of metric oscillations in the solution, with their depth depending on the amplitude of these oscillations.

By analogy with the ultra-compact star case, the presence of these bottleneck features indicates that the space-time generated in our solutions should have the ability to trap null particles. It is worth emphasizing, however that Einstein--Dirac solitons exist in a regime far removed from that of ultra-compact stars. The structures here have a radial extent of a few hundred Planck lengths, and contain fewer than 100 particles, yet are still able to produce a similar distortion of space-time, despite their much smaller scale.

\section{Fermion self-trapping}
\label{secFermSelfTrap}

Having introduced the concept of optical geometry, we now discuss its relation to the structure of Einstein--Dirac solitons, and present the main result of this paper --- the phenomenon of fermion self-trapping. 

To this end, we must analyze more thoroughly the matter component of our solutions. Consider the fermion number density:
\begin{equation}
n_f(r)=\frac{\kappa T(r)}{r^2}\left(\alpha(r)^2+\beta(r)^2 \right ),
\end{equation}
defined such that $\int n_f(r)\sqrt{-g}\,\mathrm{d}^3x=\kappa$. This quantity can be straightforwardly interpreted as the number of fermions per unit volume.

\begin{figure}
	\includegraphics[width=0.9\linewidth]{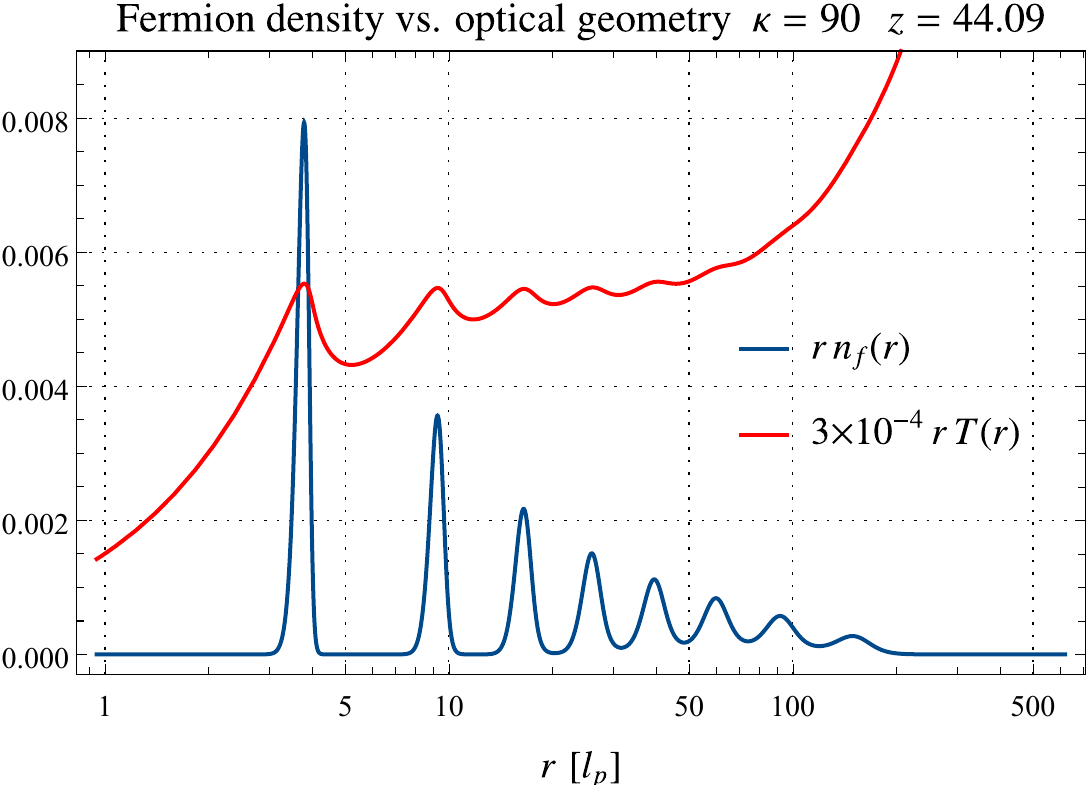}\\
	\vspace{5pt}
	\includegraphics[width=0.9\linewidth]{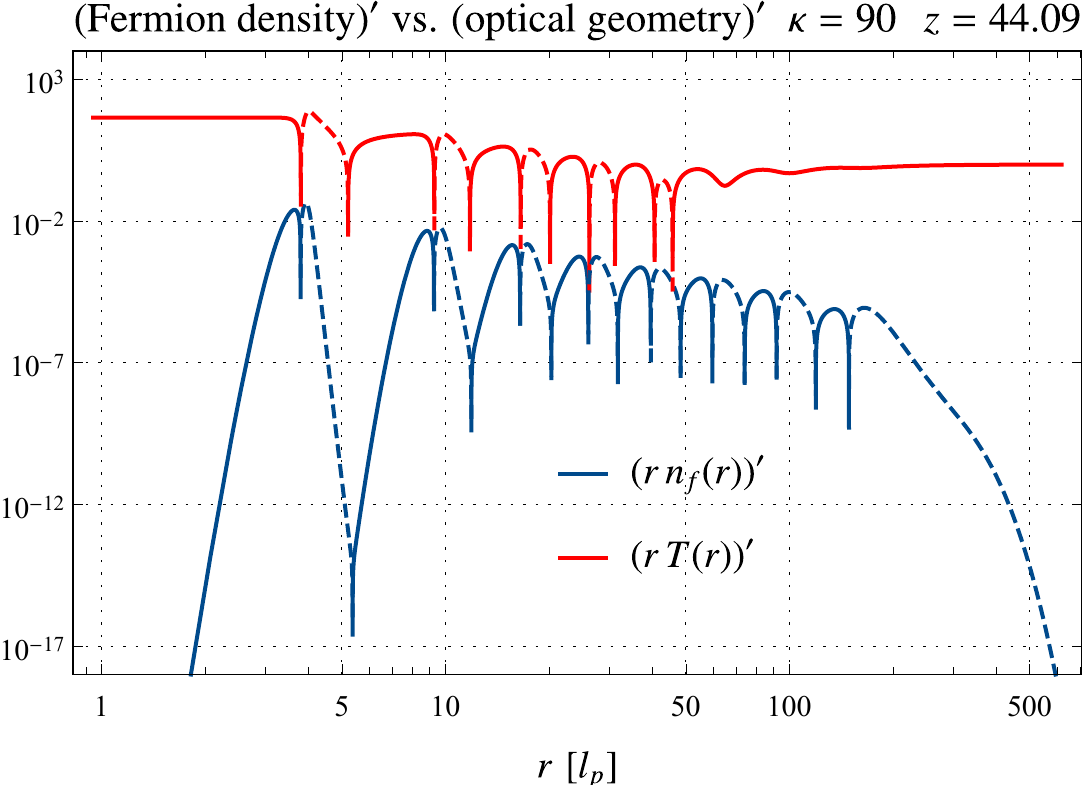}
	\caption{Plots showing (top) the fermion number density with rescaled optical geometry overlaid, and (bottom) their derivatives, for the highest redshift solution considered previously ($\kappa=90$, $z=44.09$). The dashed lines indicate where quantities become negative. Note that, at low radii, the peaks and troughs in the number density line up almost precisely with the bulges and necks of the optical geometry, suggesting that the fermions have become trapped in the bottleneck structures. This agreement breaks down at larger radii, where the fermions become less relativistic, and hence more easily trapped than a null particle. This results in additional peaks occurring in the number density despite there being no corresponding circular null geodesics.}
	\label{figNumDen}
\end{figure}

The top panel of Fig.~\ref{figNumDen} shows the fermion number density for the highest redshift solution considered previously ($\kappa=90$, $z=44.09$), onto which is superimposed the optical geometry `radial' co-ordinate $\rho=rT(r)$, the peaks and troughs of which correspond to the bulges and necks in the optical geometry.

As can be clearly seen, the fermion number density consists of a series of peaks, the radii of which (at least for the first 4 or 5 peaks) correspond to the locations of the necks in the optical geometry. This agreement can be seen more clearly in the bottom panel of Fig.~\ref{figNumDen}, which shows the derivative of the plot above. These results suggest that the fermion wavefunction is responding to the space-time in the same way as a classical null particle would respond to the optical geometry. That is to say the fermions become trapped around the stable circular null geodesics in the space-time, resulting in the number density becoming highly peaked at these points.

It is worth remembering, however, that the space-time and matter components of the system are not independent, but are determined self-consistently with respect to each other. There is no fixed background metric onto which we are adding fermions --- the space-time structure is instead created by the mass distribution and vice versa. The overall interpretation is therefore that the fermions are becoming trapped within the space-time created by their own energy density. It is in this sense that we refer to this phenomenon as fermion `self-trapping'.

We also observe that the depth of a bottleneck is related to its ability to trap null particles --- a more pronounced bottleneck results in a larger density of fermions being trapped within a narrower region. At low radii, therefore, the effect of the fermion self-trapping is so extreme that spatially well-separated shells of fermion density arise, in-between which the probability of finding a fermion is near zero. At higher radii, however, the bottlenecks are less pronounced and the peaks in the fermion wavefunction begin to merge together.

We emphasize that the appearance of this type of structure is in stark contrast to what occurs for low-redshift solutions, in which the number density contains a single peak, consistent with the expectations for a filled shell of high-angular momentum particles. In the high-redshift solutions, however, the space-time has become so distorted as to convert this single peak into something more akin to a multiple-shell model.

Why does the fermion wavefunction respond so precisely to the optical geometry? As discussed in Sec.~\ref{secOptGeom}, the optical geometry formalism applies strictly to null particles, whereas our states contain fermions with a (large) non-zero mass $m$. The answer to this is not immediately obvious. One possible explanation is related to the fact that, classically speaking, the fermions are highly relativistic in the inner regions of the soliton, and hence their trajectories should differ only slightly from those of massless particles. Furthermore, the fermions become less relativistic as radius increases, and so we would expect the match between the number density and the optical geometry to break down in the outer regions of the solution, which is precisely what we see in Fig.~\ref{figNumDen}.

Finally, we note that, strictly speaking, Fig.~\ref{figNumDen} shows plots of $r\,n_f(r)$, rather than the number density itself. For such high values of $\kappa$ as we are considering here, this distinction is fairly inconsequential, but it is interesting to note that it is indeed the former quantity which responds more precisely to the optical geometry. The reason for this is again unclear. One suggestion is that, again due to relativistic effects, the fermions should have an effective mass $\approx \omega T+m$. It is then this effective mass which would respond to the optical geometry, and since $T\sim 1/r$ within the bulk of the soliton, this can account for the additional factor of $r$ required.

\section{Binding energy and mass-radius spirals}
\label{secSpirals}

One of the more intriguing properties of Einstein--Dirac solitons, discovered initially by Finster \textit{et al.} in their original paper \cite{FSY1999original}, is the appearance of spiral structures when studying the family of solutions found by continuously varying the central redshift. Values of quantities such as the fermion mass $m$, fermion energy $\omega$ and soliton radius $R$ all exhibit oscillations as redshift is increased, resulting in spiraling behavior when plotted against each other.

In this section, we demonstrate how the structure of these spirals change when considering solutions with high values of $\kappa$, and how features of the plots can be explained by the fermion self-trapping interpretation. Note that we are able to generate solutions with much higher redshift than in previous works, and can therefore see much further within the spiral structures.

\begin{figure*}
	\includegraphics[width=0.32\linewidth]{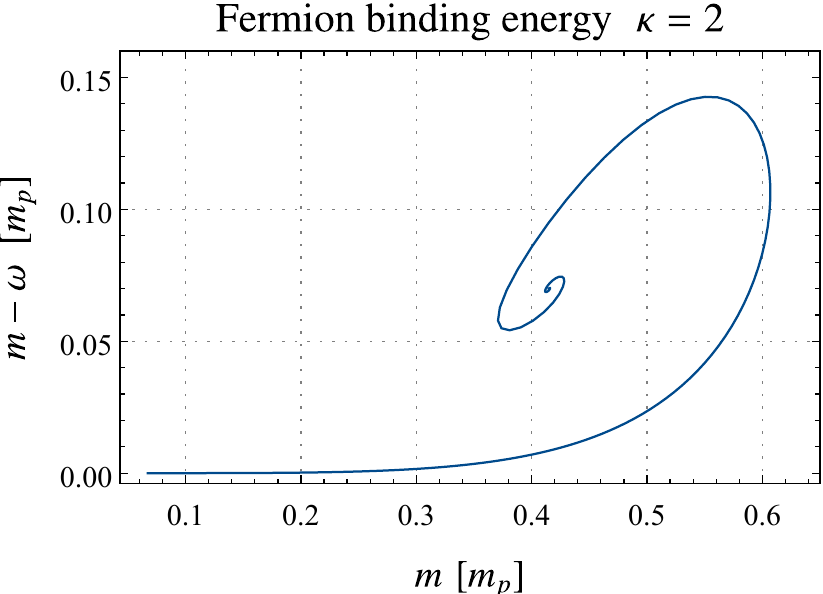}\hspace{5pt}
	\includegraphics[width=0.32\linewidth]{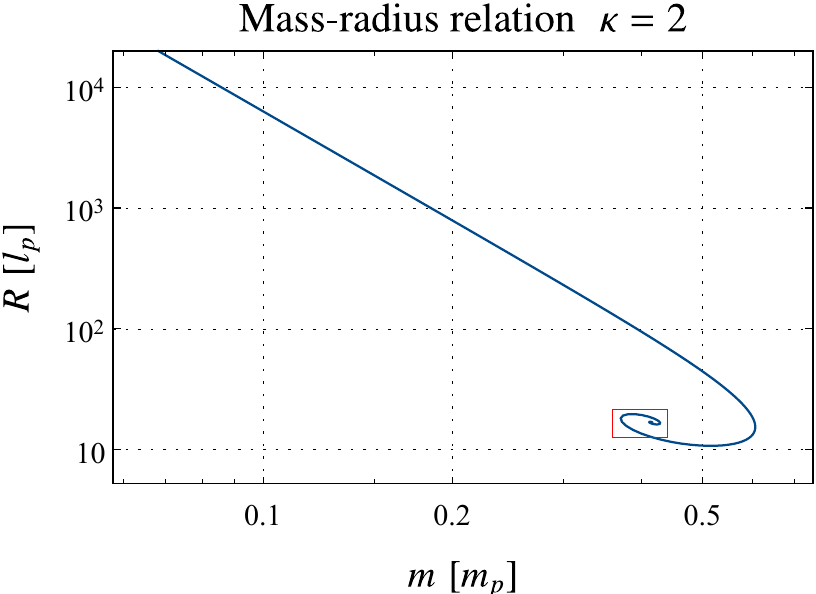}\hspace{5pt}
	\includegraphics[width=0.32\linewidth]{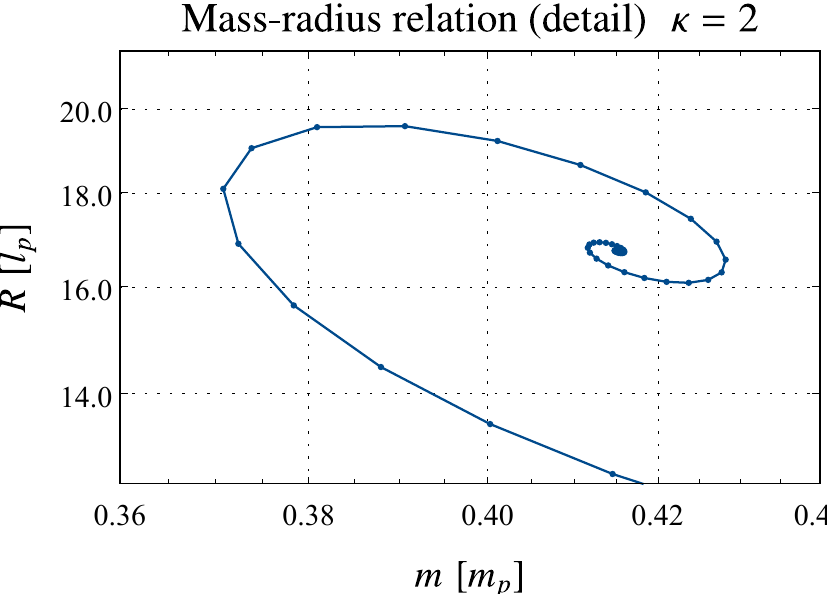}\\
	\vspace{5pt}
	\includegraphics[width=0.32\linewidth]{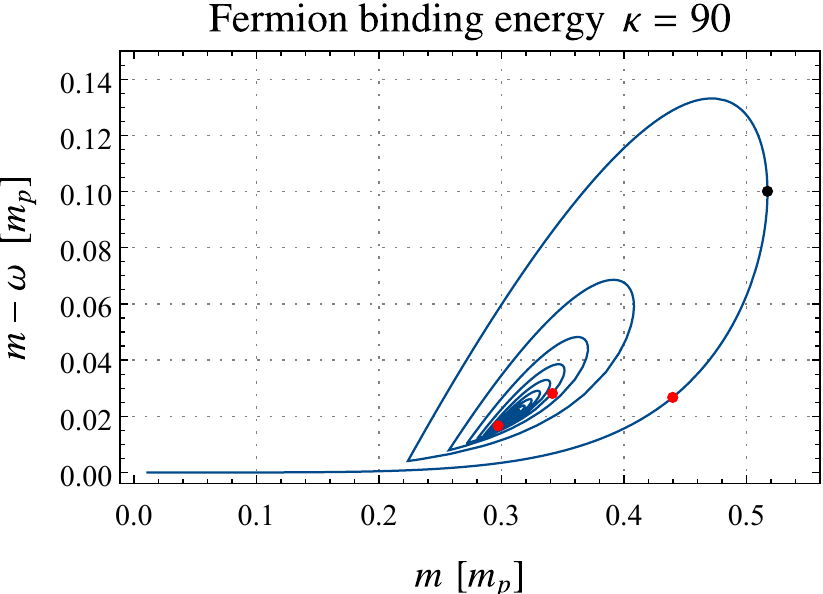}\hspace{5pt}
	\includegraphics[width=0.32\linewidth]{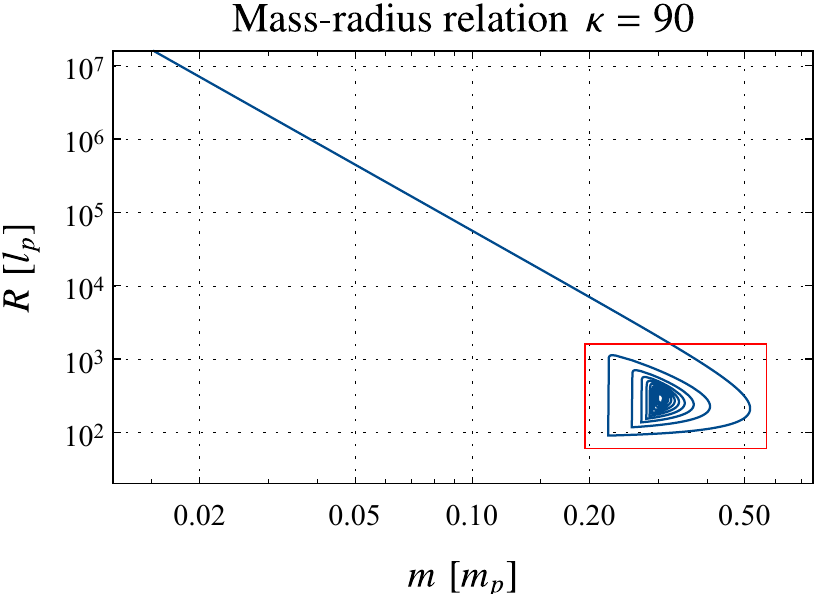}\hspace{5pt}
	\includegraphics[width=0.32\linewidth]{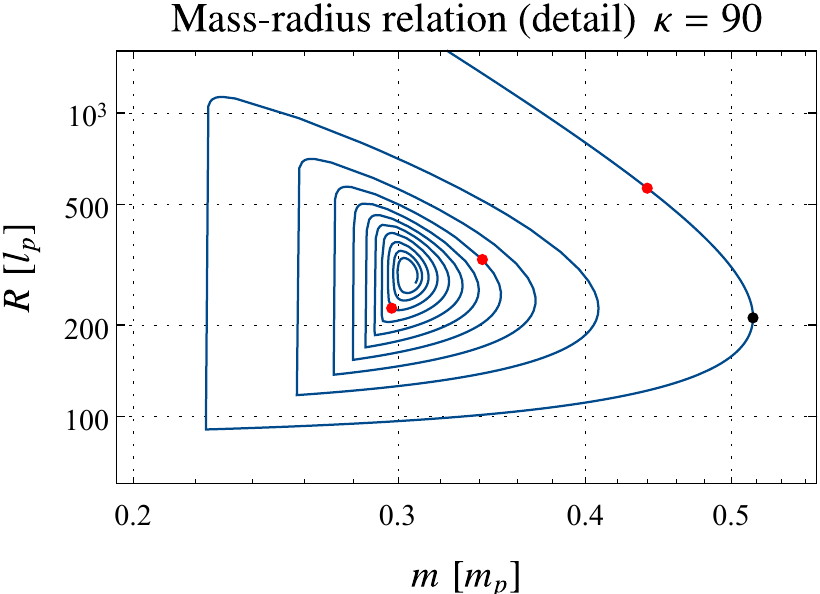}
	\caption{Fermion binding energy and mass-radius plots for states with $\kappa=2$ (top) and $\kappa=90$ (bottom). The right-hand figures show details of the regions enclosed by the red boxes in the middle plots. For both $\kappa$ values, a spiraling behavior arises as redshift is increased, although kinks and discontinuities form in the high $\kappa$ case. The black dots show the position of the stable to unstable transition point, which occurs at the point of maximum fermion mass. The red dots indicate the redshift values of the three solutions plotted in Fig.~\ref{figHighKappa}. The two higher redshift cases are expected to be unstable, as they lie beyond the transition point.}
	\label{figSpirals}
\end{figure*}

Figure \ref{figSpirals} shows spiral plots for the cases of $\kappa=2$ and $\kappa=90$. In the left-hand panels of the figure, we show the fermion binding energy $m-\omega$, as a function of fermion mass, noting that $m-\omega$ is always positive, consistent with the notion of the fermions in our solutions being bound. The $\kappa=2$ curve is a smooth spiral, leaving the origin at low central redshift, and spiraling inwards to a limiting configuration as redshift is increased. For the case of $\kappa=90$, however, this smooth curve is replaced by a function which has a much larger extent and contains a number of sharp kinks, but still retains the overall spiral structure.

Also shown in Fig.~\ref{figSpirals} are the mass-radius relations for the two $\kappa$ values, which plot the fermion mass $m$ versus the radial extent $R$ of solutions. We take $R$ to be the radius that encloses 99.9\% of the ADM mass $M$, which is defined by $M=\lim_{r\rightarrow \infty}\frac{r}{2}(1-A(r))$. At low central redshift, solutions are highly diffuse, but become more compressed as redshift is increased, with the curve ultimately spiraling inwards to a limiting, infinite redshift, configuration. For $\kappa=90$, this mass-radius relation differs significantly in structure to the $\kappa=2$ case. The spiral curve now contains discontinuities (noting that the vertical lines in the mass-radius plot contain no solutions along them), in which the radial extent of the solution increases significantly over a very small redshift range.

The appearance of these kinks and discontinuities at high $\kappa$ can be understood by considering again the fermion self-trapping effect. The overall picture is as follows. For low-redshift solutions, the fermions are arranged in a single shell, with a single peak in the fermion number density. As redshift is increased, however, the inner regions of the soliton become more compressed, and at some critical redshift, the space-time becomes distorted enough to admit a stable circular orbit at a radius beyond this single peak. It is now possible for fermions to become trapped around this region, and the fermion wavefunction therefore redistributes itself such that it is doubly peaked. This results in the radial extent of the solution increasing within a very short redshift range, thus explaining the first discontinuous jump in the mass-radius relation. The appearance of this new trapping region is also responsible for the first kink in the binding energy plot. As the redshift is increased further, subsequent trapping regions appear, each resulting in a further discontinuity/kink in the spiral curves.

Note that these jumps stop at sufficiently high redshift, and the spirals become smooth. This can be attributed to the fact that the later trapping regions which form at higher redshift are much less pronounced, and so have a lesser trapping ability, resulting in broader peaks in the fermion number density. Later stable circular orbits therefore form at radii which are within the previous trapping region, and consequently no discontinuous jump in radial extent occurs.

To close this section, we point out that spiraling behavior in mass-radius relations is known to arise in astrophysical situations, for example in theories describing neutron stars and white dwarfs. These objects have a maximum stable mass beyond which degeneracy pressure cannot prevent gravitational collapse. Similarly, Einstein--Dirac solitons exhibit a maximum fermion mass (as shown by the black dot in Fig.~\ref{figSpirals}), beyond which no static solutions exist. Unlike neutron stars or white dwarfs, however, it is not degeneracy pressure which prevents our states from collapsing, but the effects of the uncertainty principle. We find that, for $\kappa=90$, this maximum fermion mass takes the value of $0.517\,m_p$, where $m_p$ is the Planck mass. In the Appendix, we discuss how the mass-radius relations for white dwarfs and boson stars differ from Einstein--Dirac solitons, and also derive scaling relationships between quantities which hold at low redshift.

\section{Relationship between fermion energy and optical geometry}
\label{secFrequency}

In \cite{Abramowicz1997ogUltracompact} it was shown that the frequency of trapped gravitational wave modes around an ultra-compact star can be determined from properties of its optical geometry. We now demonstrate that a similar relationship exists for high-redshift Einstein--Dirac solitons, in that the fermion energy can be obtained by considering the travel time around null geodesics in the optical geometry. This relies on a WKB-type argument, in which we assume that the fermion wavefunction can be approximated by combining the classical paths of null particles, with appropriate weightings.

Consider first a classical null particle moving on a circular geodesic in the space-time of one of our high-redshift solutions. We take its path to lie in the equatorial plane $\theta=\pi/2$ without loss of generality. Its equation of motion can be derived by setting $\mathrm{d}s^2=\mathrm{d}r^2=0$ in Eq.~(\ref{metric}), resulting in
\begin{align}
0&\,=\,-\frac{\mathrm{d}t^2}{T(r)^2}+r^2\mathrm{d}\phi^2,\notag\\
\Rightarrow\;\; \frac{\mathrm{d}t}{\mathrm{d}\phi}&\,=\,rT(r).
\end{align}
Integrating over one complete orbit therefore gives an expression for the travel time $\tau_c$ around a circular null geodesic, as measured by an observer in the flat space as $r\rightarrow\infty$:
\begin{equation}
\tau_c(r)=2\pi r T(r)=2\pi \rho.
\label{circTT}
\end{equation}
Note of course that null circular orbits occur only when $(rT)'=0$, and so this relation is valid only at the specific radii of the necks and bulges in the optical geometry. Note also that this expression for the travel time can be inferred directly from the optical geometry embedding diagram --- it is simply the distance traveled around a circular orbit with `radius' $\rho=rT$.

Recall that the locations of peaks in the fermion number density correspond to the radii of bulges in the optical geometry, i.e positions of stable circular null geodesics. To a first approximation, the dominant contribution to the fermion wavefunction should therefore come from the classical orbits at these positions. The travel time around each individual bulge can be calculated from the optical geometry, using Eq.~(\ref{circTT}), and an overall mean travel time can then be obtained by taking a weighted average. In a true WKB analysis the classical action of each path would provide a natural weighting, but here we instead use the relative width of each trapping region, which has been observed to be roughly proportional to the number of fermions trapped within it.

The mean travel time in the optical geometry can therefore be expressed as
\begin{equation}
\tau_{og,circ}=\frac{\sum_n (r_n^+-r_n^-)\tau_c(r_n)}{\sum_n (r_n^+-r_n^-)},
\end{equation}
where $r_n$ is the radius of the $n^{th}$ stable null circular geodesic, and $r_n^\pm$ are the minimum and maximum radii that define the region within which a classical particle can become trapped.

In order to link the travel time around a circular null geodesic to the expected fermion energy, we make use of the following argument. Assume the classical particle is now replaced by a planar matter wave of energy $\omega_p$, propagating in the $+\phi$ direction with the form $e^{i(j\phi-\omega_p t)}$. Here, $j$ is the angular momentum of each constituent fermion, equal to $(\kappa-1)/2$ for our solutions. For constructive interference to occur, the phase acquired in one temporal period $\tau$ must equal the phase acquired in one spatial orbit, i.e. $\omega_p\tau=2\pi L$. This provides us with the following relation between energy $\omega_p$ and travel time $\tau$ around a circular orbit:
\begin{equation}
\omega_p=\frac{\kappa-1}{2}\frac{2\pi}{\tau}.
\label{freqNonRel}
\end{equation}
The above argument, however, is both non-relativistic and implicitly relies on the assumption of a flat space-time. We should not therefore expect this relationship to hold exactly in the case of our high-redshift solutions, but rather be a first approximation. Since it is unclear precisely how to modify this argument, we instead make use of the `power-law' solution detailed in \cite{Bakucz2019powerLaw}, for which an analytic relationship exists between fermion frequency and circular orbit travel time:
\begin{equation}
\omega_{pl}=\sqrt{\frac{\kappa^2}{4}-\frac{1}{3}}\,\frac{2\pi}{\tau}\equiv \frac{2\pi\xi}{\tau}.
\label{freqRel}
\end{equation}
Note that, in the limit of infinite $\kappa$, this expression agrees with that found by the non-relativistic argument above.

Using this relation, we can now obtain the following expression for the fermion energy as predicted from the travel time of circular paths in the optical geometry:
\begin{equation}
\omega_{og,circ}=\frac{2\pi\xi}{\tau_{og,circ}}.
\end{equation}

\begin{figure}
	\includegraphics[width=0.9\linewidth]{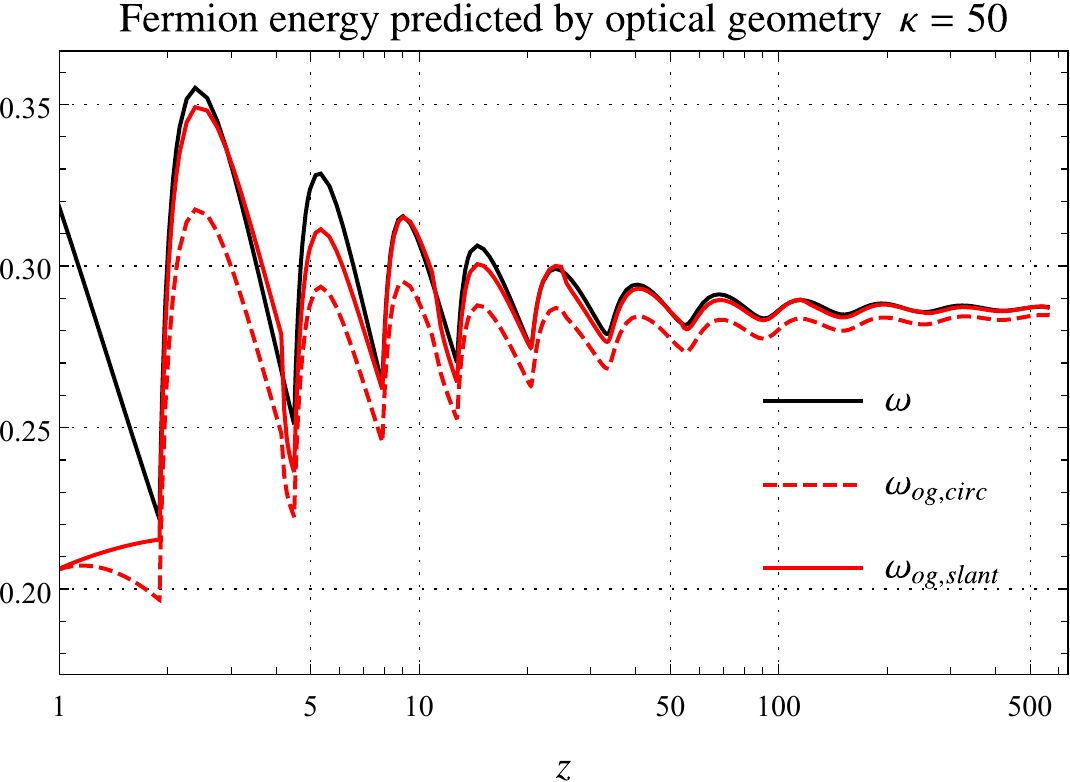}
	\caption{The true fermion frequency $\omega$ (black), plotted as a function of central redshift for the case of $\kappa=50$, alongside the predictions from the optical geometry found by using only circular orbits (dashed red) and including slanted orbits (solid red). All three curves show the expected oscillatory behavior, with each kink corresponding to the appearance of a new trapping region. The predicted frequency shows a clear improvement when slanted orbits are included. Note that the curves do not match to the extreme left of the plot as this redshift range is prior to the appearance of the first trapping region.}
	\label{figFrequency}
\end{figure}

The dashed red line in Fig.~\ref{figFrequency} shows this prediction alongside the true fermion energy $\omega$ (in black), for the case of $\kappa=50$, calculated over a range of redshift values. Our predicted frequency exhibits the same overall behavior as the true frequency, indicating a clear relationship, but the value is consistently lower than expected.

To improve the numerical agreement between the curves, we note that, although the fermion number density is heavily peaked around the bulges in the optical geometry, there is still a substantial spreading around these points. It is therefore insufficient to consider only the paths which occur precisely at the stable circular null geodesics. Classically speaking, particles can become trapped within regions around the bulges, traversing `rosette-type' orbits bounded by some minimum and maximum radius. We approximate these, somewhat crudely, as slanted circular orbits, such that their travel time can be straightforwardly calculated from the optical geometry as 
\begin{equation}
\tau_{slant}(r,r_c)=2\pi\sqrt{r^2T(r)^2+(r-r_c)^2},
\end{equation}
where $r_c$ is the radius of the stable circular obit around which the particle is trapped. An average travel time for each bulge can then be calculated by varying the value of $r$ between the limits of the trapping region. As an approximation, each path is weighted equally, although in a true WKB analysis the action would provide a natural weighting. The mean travel time of null particles trapped around a bulge in the optical geometry located at $r=r_c$ is therefore
\begin{equation}
\tau_{bulge}(r_c)=\frac{1}{(r^+-r^-)}\int_{r^-}^{r^+}\tau_{slant}(r,r_c)\,\mathrm{d}r.
\end{equation}
Now averaging over all the trapping regions in the solution, and converting to an energy using Eq.~(\ref{freqRel}), gives the following expression for the fermion energy predicted by the optical geometry, now including slanted orbits:
\begin{equation}
\omega_{og,slant}=\frac{2\pi\xi\sum_n (r_n^+-r_n^-)}{\sum_n(r_n^+-r_n^-)\tau_{bulge}(r_n)}.
\end{equation}

A plot of this quantity as a function of central redshift is shown as the solid red curve in Fig.~\ref{figFrequency}. This is clearly an improvement on the prediction obtained by considering only circular orbits, although still not an exact match to the true fermion frequency. A more thorough analysis of the problem would require a true WKB approximation, in which all paths are considered, each weighted by their respective classical action. By using the optical geometry, we are also making the implicit assumption that the fermions are massless, which is not strictly the case. Given the obvious shortcomings in our analysis, it is perhaps surprising that such good agreement between the predicted and true fermion energies can be obtained. 

\section{Discussion and Outlook}
\label{secConclusion}

We have presented solutions to the coupled Einstein--Dirac system corresponding to gravitationally localized states of fermions, focusing on the limits of high particle number and central redshift. We have shown that these solutions differ significantly from their low-redshift counterparts, with this difference being attributed to the appearance of a fermion self-trapping effect.

There are a number of important points to note. The first is that the high-redshift solutions, in which the fermion trapping is in evidence, all lie beyond the stable to unstable transition point in the binding energy curves (see Fig.~\ref{figSpirals}), and as such are expected to be dynamically unstable to infinitesimal perturbations. Given the strong gravitational effects present in these solutions, and the terminology of `trapping', one might be forgiven for expecting such states to be stable. For clarity, we emphasize that this is not the case.

We also note that, while the results presented here have been restricted to high $\kappa$ solutions, the fermion self-trapping effect is in fact present even in the case of just two fermions. Indeed at any $\kappa$, given sufficiently high redshift, circular null geodesics will occur around which the fermions can become trapped. With low numbers of particles, however, the back-reaction of the matter on the metric is weak, and so the bottlenecks in the optical geometry are relatively shallow. This results therefore in only small oscillations appearing in the fermion fields (such as those seen in \cite{Bakucz2019powerLaw}), corresponding to small over-- and under-- densities. Only when the fermion number is large does the back-reaction on the space-time become strong enough for the trapping to cause such extreme effects as the appearance of spatially separated shells.

Furthermore, we have shown that the kinks and discontinuities which appear at high $\kappa$ in the binding energy and mass-radius spirals can also be explained by the fermion self-trapping interpretation. This can be extended to low $\kappa$, where each new spiral corresponds to a new peak (however small) appearing in the fermion number density. We have also shown that the value of the fermion energy $\omega$ can be calculated purely from properties of the solution's space-time. Together, these results suggest that the appearance of the spiral structure itself may in fact be due to the fermion self-trapping effect. Since spiral structures of a similar kind are known to exist in theories describing astrophysical objects such as neutron stars, white dwarfs and boson stars, this raises the possibility that a similar self-trapping effect may be present in these scenarios.

The space-time generated in our solutions is also interesting in its own right. Single bottlenecks (arising from a pair of circular null geodesics) are well known to arise when considering compact objects, but the appearance of multiple bottlenecks is not so prevalent. Previous studies by Karlovini \textit{et al.} in \cite{Karlovini2001multipleNecks} and \cite{Karlovini2002multipleNecks} have shown that these can arise in principle, but we believe that our high-redshift Einstein-Dirac solitons constitute the first specific physical systems in which such multiple bottlenecks have been observed to occur. The reason behind their appearance is, however, currently unclear.

We emphasize that the appearance of the fermion self-trapping effect relies heavily on the implicit inclusion of back-reaction in the Einstein--Dirac system. Often the approach in semiclassical gravity is either to neglect the back-reaction, or to assume that it can be treated perturbatively, a necessary approach when quantum field theory is involved. In the Einstein--Dirac system, however, the matter is treated simply as a quantum wavefunction, allowing for the study of systems in which the effect of back-reaction is strong. Indeed, the fermion self-trapping effect discussed here is an example of a situation in which the back-reaction can dominate the behavior of the system.

We close by indicating a few possible directions in which this work could be extended.  Recall first that the fermion self-trapping effect becomes stronger as the particle number $\kappa$ is increased --- the bulges in the optical geometry become more pronounced, and the peaks in the fermion number density become progressively narrower.
This suggests that, in the limit of strictly infinite $\kappa$, the fermion wavefunction may split into a series of delta functions. It would therefore be worth investigating whether an analytic solution describing such a situation exists in the high $\kappa$ limit.

A further extension would be to consider the effect of an additional repulsive force in the system. The most obvious candidate is charge, which can be achieved by considering the Einstein--Dirac--Maxwell equations, for which particle-like solutions have previously been generated \cite{FSY1999maxwell}. Given the appearance of spiral structures also in this system, we would expect a similar fermion self-trapping effect to be present at high redshift. The addition of a repulsive force between the fermions may, however, cause the trapping peaks to broaden, perhaps preventing the formation of similar multiple-shell-like solutions to those shown here.

Finally, we point out that the dynamical time-evolution of Einstein--Dirac solitons has not as yet been fully explored. Although a stable branch of solutions is known to exist \cite{FSY1999original}, high-redshift solutions, including those exhibiting the fermion self-trapping effect, are expected to be unstable. The issue of the precise behavior of unstable solutions will be addressed in a future publication. Of particular relevance to the discussion here would be to determine the impact of the self-trapping effect on the time dynamics of high-redshift solutions.

\appendix*

\section{Low redshift relationships}
\label{appLowRedshift}
At low central redshift, the spatial extent of an Einstein--Dirac soliton is such that relativistic effects are negligible, and the Einstein--Dirac equations reduce to their non-relativistic counterpart, the Newton--Schrödinger system (see \cite{Stuart2010newtonianLimit} and \cite{Giulini2012schrodingerNewton} for details). In this low-redshift regime, one can obtain analytic scaling relations which hold between certain properties of the solutions.

\subsection{Mass-radius relations}
We present first a derivation of the relationship between the ADM mass $M$ and radial extent $R$ of low-redshift Einstein--Dirac solitons. We also include similar derivations valid for neutrons stars/white dwarfs, and boson stars, to highlight the differences.

\subsubsection{Einstein--Dirac solitons}
Recall that spiral structures arise when considering the mass-radius relationships for Einstein--Dirac solitons (see Fig.~\ref{figSpirals}). Similar curves are obtained if instead the ADM mass $M$ is plotted against $R$. Regardless of the value of $\kappa$, the low-redshift portions of these curves (in which the fermion mass is small and solutions therefore have a large radial extent), are found to very well approximate $M\sim R^{-1/3}$. We outline below an analytic derivation of this relationship.

In this low-redshift regime, localized states exist under the balance between the Newtonian gravitational attraction and the kinetic energy of the fermions. Equating the total non-relativistic kinetic and gravitational energies for a system of $\kappa$ particles gives:
\begin{equation}
\kappa\frac{p^2}{2m}\approx\kappa\frac{GMm}{R},
\end{equation}
where $p$ is the momentum of each constituent fermion. 

To satisfy the uncertainty principle, we require $\Delta x \Delta p \sim 1$ for each individual fermion wavefunction. Since our fermions are arranged in a filled shell, they do not experience the exclusion principle, and so each fermion has an effective volume proportional to $R^3$, implying $\Delta x\sim R$. It follows that $p\sim 1/R$, and so
\begin{equation}
\frac{1}{2mR^2}\sim\frac{GMm}{R}.
\end{equation}
Since the fermion mass $m$ is not constant along the mass-radius curve, it must be eliminated in favor of the ADM mass $M$. In the non-relativistic limit, $M\approx\kappa m$, giving
\begin{align}
\frac{\kappa}{2MR^2}&\sim \frac{GM^2}{\kappa R}, \notag \\
\Rightarrow\;\;\frac{1}{M^3}&\sim \frac{2GR}{\kappa^2}.
\end{align}
The relationship $M\sim R^{-1/3}$ therefore holds, for constant $\kappa$, in the low-redshift limit.

\subsubsection{White Dwarfs / Neutron Stars}
We note that this relationship differs from the well-known expression valid for low-mass astrophysical fermionic objects, such as white dwarfs and neutron stars, for which $M\sim R^{-3}$. To highlight the difference, we perform the analogous calculation for objects of this type.
Taking the white dwarf/neutron star to consist of $N$ fermions of mass $m$, balance is as before between kinetic and gravitational energy:
\begin{equation}
N\frac{p^2}{2m}\approx N\frac{GMm}{R}.
\end{equation}
Now, however, there is the extra effect of degeneracy pressure to take into account. This implies that each fermion takes up an effective volume proportional to $R^3/N$, from which the uncertainty principle implies $p\sim N^{1/3}R^{-1}$. Hence
\begin{equation}
\frac{N^{2/3}}{2mR^2}\sim \frac{GMm}{R}.
\end{equation}
When considering astrophysical objects, the fermion mass is taken to be the electron mass and is hence fixed. It is therefore the number of particles which now varies along the mass-radius curve, and so $N$ must be eliminated in favor of $m$. As before, $M\approx Nm$, and so
\begin{align}
\frac{M^{2/3}}{2m^{5/3}R^2}&\sim \frac{GMm}{R},\notag\\
\Rightarrow\:\:\frac{1}{M^{1/3}}&\sim m^{8/3}R.
\end{align}
This recovers the usual $M\sim R^{-3}$ relationship.

\subsubsection{Boson Stars}
For completeness, we also derive the expected mass-radius relation for boson stars with low central densities, for which $M\sim R^{-1}$. Since we are dealing with bosons, there is no degeneracy pressure, with each boson taking up an effective volume $\propto R^3$, implying $p\sim 1/R$. As for the Einstein--Dirac case, this leads to
\begin{equation}
\frac{1}{2mR^2}\sim\frac{GMm}{R}.
\end{equation}
Mass-radius relations, such as those found in \cite{Seidel1990bsDynamical} and \cite{Schunck2003bsGeneralRelativistic}, are then generated by varying the number of bosons $N$, treating the boson mass $m$ as a constant. The above relationship can therefore be directly rearranged to show $M\sim R^{-1}$, which should hold in the low central density limit.

\subsection{Explicit redshift relationships}
Returning to the case of Einstein--Dirac solitons, we now derive analytic expressions for how the fermion energy $\omega$, fermion mass $m$ and soliton radius $R$ scale explicitly with central redshift $z$, in the low-redshift limit.

At low redshift, solutions are non-relativistic and space-time is approximately flat, i.e. $A(r),T(r)\approx 1$. We note that the metric field $T(r)$ deviates only slightly from its central value, $T(0)=1+z$, throughout the matter bulk, before latching on to the Schwarzschild solution
\begin{equation}
T_{sch}(r)=(1-2GM/r)^{-1/2}\approx 1+GM/r,
\end{equation}
at approximately the radius of the soliton. We can therefore identify $z\sim GM/R$. Using this, along with the mass-radius relation $M\sim R^{-1/3}$ derived previously, and noting that $M\approx \kappa m$, we can directly infer the following scaling relations:
\begin{align}
R&\sim z^{-3/4}\,;\\
M&\sim z^{1/4}\,;\\
m&\sim z^{1/4}\,.
\end{align}

To include the fermion energy $\omega$ in this argument requires information about the ground state solution of the Newton-Schrödinger system. This is analogous to that of the Bohr model of the hydrogen atom, with the electrostatic attraction replaced by gravity i.e.
\begin{equation}
\frac{e^2}{4\pi\epsilon_0}\rightarrow GMm.
\end{equation}
Analogous quantities to the hydrogen atom Bohr radius $a_0$ and Rydberg constant $R_H$ can then be written as:
\begin{align}
a_0&=\frac{4\pi\epsilon_0}{me^2}\rightarrow \frac{1}{GMm^2}\,;\\
R_H&=\frac{1}{ma_0^2}\rightarrow G^2M^2m^3.
\end{align}
Since the ground state energy $E_0$ is proportional to $R_H$, and recalling that $M\approx \kappa m$, we obtain the relation $E_0\sim m^5\sim z^{5/4}$. Identifying this ground state energy with either the fermion binding energy $m-\omega$ or the (negative) soliton binding energy $\kappa m-M$, we obtain the following scaling relations:
\begin{align}
(m-\omega) &\sim z^{5/4}\,;\\
(\kappa m-M) &\sim z^{5/4}.
\end{align}
For this to hold, we require $m$ and $\omega$ to scale in the same way, giving us our final scaling relation:
\begin{equation}
\omega \sim z^{1/4}.
\end{equation}


\begin{thebibliography}{22}%
	\makeatletter
	\providecommand \@ifxundefined [1]{%
		\@ifx{#1\undefined}
	}%
	\providecommand \@ifnum [1]{%
		\ifnum #1\expandafter \@firstoftwo
		\else \expandafter \@secondoftwo
		\fi
	}%
	\providecommand \@ifx [1]{%
		\ifx #1\expandafter \@firstoftwo
		\else \expandafter \@secondoftwo
		\fi
	}%
	\providecommand \natexlab [1]{#1}%
	\providecommand \enquote  [1]{``#1''}%
	\providecommand \bibnamefont  [1]{#1}%
	\providecommand \bibfnamefont [1]{#1}%
	\providecommand \citenamefont [1]{#1}%
	\providecommand \href@noop [0]{\@secondoftwo}%
	\providecommand \href [0]{\begingroup \@sanitize@url \@href}%
	\providecommand \@href[1]{\@@startlink{#1}\@@href}%
	\providecommand \@@href[1]{\endgroup#1\@@endlink}%
	\providecommand \@sanitize@url [0]{\catcode `\\12\catcode `\$12\catcode
		`\&12\catcode `\#12\catcode `\^12\catcode `\_12\catcode `\%12\relax}%
	\providecommand \@@startlink[1]{}%
	\providecommand \@@endlink[0]{}%
	\providecommand \url  [0]{\begingroup\@sanitize@url \@url }%
	\providecommand \@url [1]{\endgroup\@href {#1}{\urlprefix }}%
	\providecommand \urlprefix  [0]{URL }%
	\providecommand \Eprint [0]{\href }%
	\providecommand \doibase [0]{https://doi.org/}%
	\providecommand \selectlanguage [0]{\@gobble}%
	\providecommand \bibinfo  [0]{\@secondoftwo}%
	\providecommand \bibfield  [0]{\@secondoftwo}%
	\providecommand \translation [1]{[#1]}%
	\providecommand \BibitemOpen [0]{}%
	\providecommand \bibitemStop [0]{}%
	\providecommand \bibitemNoStop [0]{.\EOS\space}%
	\providecommand \EOS [0]{\spacefactor3000\relax}%
	\providecommand \BibitemShut  [1]{\csname bibitem#1\endcsname}%
	\let\auto@bib@innerbib\@empty
	%</preamble>
	\bibitem [{\citenamefont {Lee}\ and\ \citenamefont
		{Pang}(1987)}]{Lee1987solitonStars}%
	\BibitemOpen
	\bibfield  {author} {\bibinfo {author} {\bibfnamefont {T.~D.}\ \bibnamefont
			{Lee}}\ and\ \bibinfo {author} {\bibfnamefont {Y.}~\bibnamefont {Pang}},\
	}\bibfield  {title} {\bibinfo {title} {Fermion soliton stars and black
			holes},\ }\href@noop {} {\bibfield  {journal} {\bibinfo  {journal} {Phys.
				Rev. D}\ }\textbf {\bibinfo {volume} {35}},\ \bibinfo {pages} {3678}
		(\bibinfo {year} {1987})}\BibitemShut {NoStop}%
	\bibitem [{\citenamefont {Finster}\ \emph
		{et~al.}(1999{\natexlab{a}})\citenamefont {Finster}, \citenamefont
		{Smoller},\ and\ \citenamefont {Yau}}]{FSY1999original}%
	\BibitemOpen
	\bibfield  {author} {\bibinfo {author} {\bibfnamefont {F.}~\bibnamefont
			{Finster}}, \bibinfo {author} {\bibfnamefont {J.}~\bibnamefont {Smoller}},\
		and\ \bibinfo {author} {\bibfnamefont {S.-T.}\ \bibnamefont {Yau}},\
	}\bibfield  {title} {\bibinfo {title} {Particlelike solutions of the
			{E}instein-{D}irac equations},\ }\href@noop {} {\bibfield  {journal}
		{\bibinfo  {journal} {Phys. Rev. D}\ }\textbf {\bibinfo {volume} {59}},\
		\bibinfo {pages} {104020} (\bibinfo {year} {1999}{\natexlab{a}})}\BibitemShut
	{NoStop}%
	\bibitem [{\citenamefont {Finster}\ \emph
		{et~al.}(1999{\natexlab{b}})\citenamefont {Finster}, \citenamefont
		{Smoller},\ and\ \citenamefont {Yau}}]{FSY1999maxwell}%
	\BibitemOpen
	\bibfield  {author} {\bibinfo {author} {\bibfnamefont {F.}~\bibnamefont
			{Finster}}, \bibinfo {author} {\bibfnamefont {J.}~\bibnamefont {Smoller}},\
		and\ \bibinfo {author} {\bibfnamefont {S.-T.}\ \bibnamefont {Yau}},\
	}\bibfield  {title} {\bibinfo {title} {Particle-like solutions of the
			{Einstein}--{D}irac--{M}axwell equations},\ }\href@noop {} {\bibfield
		{journal} {\bibinfo  {journal} {Phys. Lett. A}\ }\textbf {\bibinfo {volume}
			{259}},\ \bibinfo {pages} {431} (\bibinfo {year}
		{1999}{\natexlab{b}})}\BibitemShut {NoStop}%
	\bibitem [{\citenamefont {Finster}\ \emph
		{et~al.}(2000{\natexlab{a}})\citenamefont {Finster}, \citenamefont
		{Smoller},\ and\ \citenamefont {Yau}}]{FSY2000nonAbelianBound}%
	\BibitemOpen
	\bibfield  {author} {\bibinfo {author} {\bibfnamefont {F.}~\bibnamefont
			{Finster}}, \bibinfo {author} {\bibfnamefont {J.}~\bibnamefont {Smoller}},\
		and\ \bibinfo {author} {\bibfnamefont {S.-T.}\ \bibnamefont {Yau}},\
	}\bibfield  {title} {\bibinfo {title} {The interaction of {D}irac particles
			with non-abelian gauge fields and gravity--bound states},\ }\href@noop {}
	{\bibfield  {journal} {\bibinfo  {journal} {Nucl. Phys. B}\ }\textbf
		{\bibinfo {volume} {584}},\ \bibinfo {pages} {387} (\bibinfo {year}
		{2000}{\natexlab{a}})}\BibitemShut {NoStop}%
	\bibitem [{\citenamefont {Finster}\ \emph
		{et~al.}(2000{\natexlab{b}})\citenamefont {Finster}, \citenamefont
		{Smoller},\ and\ \citenamefont {Yau}}]{FSY2000bhEinsteinDirac}%
	\BibitemOpen
	\bibfield  {author} {\bibinfo {author} {\bibfnamefont {F.}~\bibnamefont
			{Finster}}, \bibinfo {author} {\bibfnamefont {J.}~\bibnamefont {Smoller}},\
		and\ \bibinfo {author} {\bibfnamefont {S.-T.}\ \bibnamefont {Yau}},\
	}\bibfield  {title} {\bibinfo {title} {Non-existence of time-periodic
			solutions of the {D}irac equation in a {R}eissner-{N}ordstr{\"o}m black hole
			background},\ }\href@noop {} {\bibfield  {journal} {\bibinfo  {journal} {J.
				Math. Phys.}\ }\textbf {\bibinfo {volume} {41}},\ \bibinfo {pages} {2173}
		(\bibinfo {year} {2000}{\natexlab{b}})}\BibitemShut {NoStop}%
	\bibitem [{\citenamefont {Finster}\ \emph
		{et~al.}(1999{\natexlab{c}})\citenamefont {Finster}, \citenamefont
		{Smoller},\ and\ \citenamefont {Yau}}]{FSY1999bhEDM}%
	\BibitemOpen
	\bibfield  {author} {\bibinfo {author} {\bibfnamefont {F.}~\bibnamefont
			{Finster}}, \bibinfo {author} {\bibfnamefont {J.}~\bibnamefont {Smoller}},\
		and\ \bibinfo {author} {\bibfnamefont {S.-T.}\ \bibnamefont {Yau}},\
	}\bibfield  {title} {\bibinfo {title} {Non-{E}xistence of {B}lack {H}ole
			{S}olutions for a {S}pherically {S}ymmetric, {S}tatic
			{E}instein--{D}irac--{M}axwell {S}ystem},\ }\href@noop {} {\bibfield
		{journal} {\bibinfo  {journal} {Comm. Math. Phys.}\ }\textbf {\bibinfo
			{volume} {205}},\ \bibinfo {pages} {249} (\bibinfo {year}
		{1999}{\natexlab{c}})}\BibitemShut {NoStop}%
	\bibitem [{\citenamefont {Finster}\ \emph
		{et~al.}(1999{\natexlab{d}})\citenamefont {Finster}, \citenamefont {Yau},\
		and\ \citenamefont {Smoller}}]{FSY1999bhNonAbelian}%
	\BibitemOpen
	\bibfield  {author} {\bibinfo {author} {\bibfnamefont {F.}~\bibnamefont
			{Finster}}, \bibinfo {author} {\bibfnamefont {S.-T.}\ \bibnamefont {Yau}},\
		and\ \bibinfo {author} {\bibfnamefont {J.}~\bibnamefont {Smoller}},\
	}\bibfield  {title} {\bibinfo {title} {The interaction of {D}irac particles
			with non-abelian gauge fields and gravity---black holes},\ }\href@noop {}
	{\bibfield  {journal} {\bibinfo  {journal} {Mich. Math. J.}\ }\textbf
		{\bibinfo {volume} {47}},\ \bibinfo {pages} {199} (\bibinfo {year}
		{1999}{\natexlab{d}})}\BibitemShut {NoStop}%
	\bibitem [{\citenamefont {Bernard}(2006)}]{Bernard2006bh}%
	\BibitemOpen
	\bibfield  {author} {\bibinfo {author} {\bibfnamefont {Y.}~\bibnamefont
			{Bernard}},\ }\bibfield  {title} {\bibinfo {title} {Non-existence of
			black-hole solutions for the electroweak {E}instein--{D}irac--{Y}ang/{M}ills
			equations},\ }\href@noop {} {\bibfield  {journal} {\bibinfo  {journal}
			{Class. Quant. Grav.}\ }\textbf {\bibinfo {volume} {23}},\ \bibinfo {pages}
		{4433} (\bibinfo {year} {2006})}\BibitemShut {NoStop}%
	\bibitem [{\citenamefont {Herdeiro}\ \emph {et~al.}(2017)\citenamefont
		{Herdeiro}, \citenamefont {Pombo},\ and\ \citenamefont
		{Radu}}]{Herdeiro2017bosonDiracProca}%
	\BibitemOpen
	\bibfield  {author} {\bibinfo {author} {\bibfnamefont {C.~A.~R.}\
			\bibnamefont {Herdeiro}}, \bibinfo {author} {\bibfnamefont {A.~M.}\
			\bibnamefont {Pombo}},\ and\ \bibinfo {author} {\bibfnamefont
			{E.}~\bibnamefont {Radu}},\ }\bibfield  {title} {\bibinfo {title}
		{Asymptotically flat scalar, {D}irac and {P}roca stars: {D}iscrete vs.
			continuous families of solutions},\ }\href@noop {} {\bibfield  {journal}
		{\bibinfo  {journal} {Phys. Lett. B}\ }\textbf {\bibinfo {volume} {773}},\
		\bibinfo {pages} {654} (\bibinfo {year} {2017})}\BibitemShut {NoStop}%
	\bibitem [{\citenamefont {Herdeiro}\ \emph {et~al.}(2019)\citenamefont
		{Herdeiro}, \citenamefont {Perapechka}, \citenamefont {Radu},\ and\
		\citenamefont {Shnir}}]{Herdeiro2019bosonDiracProcaSpinning}%
	\BibitemOpen
	\bibfield  {author} {\bibinfo {author} {\bibfnamefont {C.}~\bibnamefont
			{Herdeiro}}, \bibinfo {author} {\bibfnamefont {P.}~\bibnamefont
			{Perapechka}}, \bibinfo {author} {\bibfnamefont {E.}~\bibnamefont {Radu}},\
		and\ \bibinfo {author} {\bibfnamefont {Y.}~\bibnamefont {Shnir}},\ }\bibfield
	{title} {\bibinfo {title} {Asymptotically flat spinning scalar, {D}irac and
			{P}roca stars},\ }\href@noop {} {\bibfield  {journal} {\bibinfo  {journal}
			{Phys. Lett. B}\ }\textbf {\bibinfo {volume} {797}},\ \bibinfo {pages}
		{134845} (\bibinfo {year} {2019})}\BibitemShut {NoStop}%
	\bibitem [{\citenamefont {Daka}\ \emph {et~al.}(2019)\citenamefont {Daka},
		\citenamefont {Phan},\ and\ \citenamefont
		{Kain}}]{Daka2019diracStarEvolution}%
	\BibitemOpen
	\bibfield  {author} {\bibinfo {author} {\bibfnamefont {E.}~\bibnamefont
			{Daka}}, \bibinfo {author} {\bibfnamefont {N.~N.}\ \bibnamefont {Phan}},\
		and\ \bibinfo {author} {\bibfnamefont {B.}~\bibnamefont {Kain}},\ }\bibfield
	{title} {\bibinfo {title} {Perturbing the ground state of {D}irac stars},\
	}\href@noop {} {\bibfield  {journal} {\bibinfo  {journal} {Phys. Rev. D}\
		}\textbf {\bibinfo {volume} {100}},\ \bibinfo {pages} {084042} (\bibinfo
		{year} {2019})}\BibitemShut {NoStop}%
	\bibitem [{\citenamefont {Bakucz~Canário}\ \emph {et~al.}(tion)\citenamefont
		{Bakucz~Canário}, \citenamefont {Lloyd}, \citenamefont {Horne},\ and\
		\citenamefont {Hooley}}]{Bakucz2019powerLaw}%
	\BibitemOpen
	\bibfield  {author} {\bibinfo {author} {\bibfnamefont {D.}~\bibnamefont
			{Bakucz~Canário}}, \bibinfo {author} {\bibfnamefont {S.}~\bibnamefont
			{Lloyd}}, \bibinfo {author} {\bibfnamefont {K.}~\bibnamefont {Horne}},\ and\
		\bibinfo {author} {\bibfnamefont {C.}~\bibnamefont {Hooley}},\ }\bibfield
	{title} {\bibinfo {title} {Infinite-red-shift bound states of {D}irac
			fermions under {E}insteinian gravity}} (\bibinfo {year} {2020, in
		preparation})\BibitemShut {NoStop}%
	\bibitem [{\citenamefont {Abramowicz}\ \emph {et~al.}(1988)\citenamefont
		{Abramowicz}, \citenamefont {Carter},\ and\ \citenamefont
		{Lasota}}]{Abramowicz1988og}%
	\BibitemOpen
	\bibfield  {author} {\bibinfo {author} {\bibfnamefont {M.~A.}\ \bibnamefont
			{Abramowicz}}, \bibinfo {author} {\bibfnamefont {B.}~\bibnamefont {Carter}},\
		and\ \bibinfo {author} {\bibfnamefont {J.-P.}\ \bibnamefont {Lasota}},\
	}\bibfield  {title} {\bibinfo {title} {Optical {r}eference {g}eometry for
			{s}tationary and {s}tatic {d}ynamics},\ }\href@noop {} {\bibfield  {journal}
		{\bibinfo  {journal} {Gen. Relativ. Gravit.}\ }\textbf {\bibinfo {volume}
			{20}},\ \bibinfo {pages} {1173} (\bibinfo {year} {1988})}\BibitemShut
	{NoStop}%
	\bibitem [{\citenamefont {Abramowicz}\ \emph {et~al.}(1997)\citenamefont
		{Abramowicz}, \citenamefont {Andersson}, \citenamefont {Bruni}, \citenamefont
		{Ghosh},\ and\ \citenamefont {Sonego}}]{Abramowicz1997ogUltracompact}%
	\BibitemOpen
	\bibfield  {author} {\bibinfo {author} {\bibfnamefont {M.~A.}\ \bibnamefont
			{Abramowicz}}, \bibinfo {author} {\bibfnamefont {N.}~\bibnamefont
			{Andersson}}, \bibinfo {author} {\bibfnamefont {M.}~\bibnamefont {Bruni}},
		\bibinfo {author} {\bibfnamefont {P.}~\bibnamefont {Ghosh}},\ and\ \bibinfo
		{author} {\bibfnamefont {S.}~\bibnamefont {Sonego}},\ }\bibfield  {title}
	{\bibinfo {title} {Gravitational waves from ultracompact stars: the optical
			geometry view of trapped modes},\ }\href@noop {} {\bibfield  {journal}
		{\bibinfo  {journal} {Class. Quant. Grav.}\ }\textbf {\bibinfo {volume}
			{14}},\ \bibinfo {pages} {L189} (\bibinfo {year} {1997})}\BibitemShut
	{NoStop}%
	\bibitem [{\citenamefont {Rosquist}(1999)}]{Rosquist1999ogTrapping}%
	\BibitemOpen
	\bibfield  {author} {\bibinfo {author} {\bibfnamefont {K.}~\bibnamefont
			{Rosquist}},\ }\bibfield  {title} {\bibinfo {title} {Trapped gravitational
			wave modes in stars with ${R}>3{M}$},\ }\href@noop {} {\bibfield  {journal}
		{\bibinfo  {journal} {Phys. Rev. D}\ }\textbf {\bibinfo {volume} {59}},\
		\bibinfo {pages} {044022} (\bibinfo {year} {1999})}\BibitemShut {NoStop}%
	\bibitem [{\citenamefont {Abramowicz}(1993)}]{Abramowicz1993bh}%
	\BibitemOpen
	\bibfield  {author} {\bibinfo {author} {\bibfnamefont {M.~A.}\ \bibnamefont
			{Abramowicz}},\ }\bibfield  {title} {\bibinfo {title} {Black holes and the
			centrifugal force paradox},\ }\href@noop {} {\bibfield  {journal} {\bibinfo
			{journal} {Sci. Am.}\ }\textbf {\bibinfo {volume} {268}},\ \bibinfo {pages}
		{74} (\bibinfo {year} {1993})}\BibitemShut {NoStop}%
	\bibitem [{\citenamefont {Karlovini}\ \emph {et~al.}(2001)\citenamefont
		{Karlovini}, \citenamefont {Rosquist},\ and\ \citenamefont
		{Samuelsson}}]{Karlovini2001multipleNecks}%
	\BibitemOpen
	\bibfield  {author} {\bibinfo {author} {\bibfnamefont {M.}~\bibnamefont
			{Karlovini}}, \bibinfo {author} {\bibfnamefont {K.}~\bibnamefont
			{Rosquist}},\ and\ \bibinfo {author} {\bibfnamefont {L.}~\bibnamefont
			{Samuelsson}},\ }\bibfield  {title} {\bibinfo {title} {Constructing stellar
			objects with multiple necks},\ }\href@noop {} {\bibfield  {journal} {\bibinfo
			{journal} {Class. Quant. Grav.}\ }\textbf {\bibinfo {volume} {18}},\
		\bibinfo {pages} {817} (\bibinfo {year} {2001})}\BibitemShut {NoStop}%
	\bibitem [{\citenamefont {Karlovini}\ \emph {et~al.}(2002)\citenamefont
		{Karlovini}, \citenamefont {Rosquist},\ and\ \citenamefont
		{Samuelsson}}]{Karlovini2002multipleNecks}%
	\BibitemOpen
	\bibfield  {author} {\bibinfo {author} {\bibfnamefont {M.}~\bibnamefont
			{Karlovini}}, \bibinfo {author} {\bibfnamefont {K.}~\bibnamefont
			{Rosquist}},\ and\ \bibinfo {author} {\bibfnamefont {L.}~\bibnamefont
			{Samuelsson}},\ }\bibfield  {title} {\bibinfo {title} {Ultracompact stars
			with multiple necks},\ }\href@noop {} {\bibfield  {journal} {\bibinfo
			{journal} {Mod. Phys. Lett. A}\ }\textbf {\bibinfo {volume} {17}},\ \bibinfo
		{pages} {197} (\bibinfo {year} {2002})}\BibitemShut {NoStop}%
	\bibitem [{\citenamefont {Stuart}(2010)}]{Stuart2010newtonianLimit}%
	\BibitemOpen
	\bibfield  {author} {\bibinfo {author} {\bibfnamefont {D.}~\bibnamefont
			{Stuart}},\ }\bibfield  {title} {\bibinfo {title} {Existence and {N}ewtonian
			limit of nonlinear bound states in the {E}instein--{D}irac system},\
	}\href@noop {} {\bibfield  {journal} {\bibinfo  {journal} {J. Math. Phys.}\
		}\textbf {\bibinfo {volume} {51}},\ \bibinfo {pages} {032501} (\bibinfo
		{year} {2010})}\BibitemShut {NoStop}%
	\bibitem [{\citenamefont {Giulini}\ and\ \citenamefont
		{Gro{\ss}ardt}(2012)}]{Giulini2012schrodingerNewton}%
	\BibitemOpen
	\bibfield  {author} {\bibinfo {author} {\bibfnamefont {D.}~\bibnamefont
			{Giulini}}\ and\ \bibinfo {author} {\bibfnamefont {A.}~\bibnamefont
			{Gro{\ss}ardt}},\ }\bibfield  {title} {\bibinfo {title} {The
			{S}chr{\"o}dinger--{N}ewton equation as a non-relativistic limit of
			self-gravitating {K}lein--{G}ordon and {D}irac fields},\ }\href@noop {}
	{\bibfield  {journal} {\bibinfo  {journal} {Class. Quant. Grav.}\ }\textbf
		{\bibinfo {volume} {29}},\ \bibinfo {pages} {215010} (\bibinfo {year}
		{2012})}\BibitemShut {NoStop}%
	\bibitem [{\citenamefont {Seidel}\ and\ \citenamefont
		{Suen}(1990)}]{Seidel1990bsDynamical}%
	\BibitemOpen
	\bibfield  {author} {\bibinfo {author} {\bibfnamefont {E.}~\bibnamefont
			{Seidel}}\ and\ \bibinfo {author} {\bibfnamefont {W.-M.}\ \bibnamefont
			{Suen}},\ }\bibfield  {title} {\bibinfo {title} {Dynamical evolution of boson
			stars: {P}erturbing the ground state},\ }\href@noop {} {\bibfield  {journal}
		{\bibinfo  {journal} {Phys. Rev. D}\ }\textbf {\bibinfo {volume} {42}},\
		\bibinfo {pages} {384} (\bibinfo {year} {1990})}\BibitemShut {NoStop}%
	\bibitem [{\citenamefont {Schunck}\ and\ \citenamefont
		{Mielke}(2003)}]{Schunck2003bsGeneralRelativistic}%
	\BibitemOpen
	\bibfield  {author} {\bibinfo {author} {\bibfnamefont {F.~E.}\ \bibnamefont
			{Schunck}}\ and\ \bibinfo {author} {\bibfnamefont {E.~W.}\ \bibnamefont
			{Mielke}},\ }\bibfield  {title} {\bibinfo {title} {General relativistic boson
			stars},\ }\href@noop {} {\bibfield  {journal} {\bibinfo  {journal} {Class.
				Quant. Grav.}\ }\textbf {\bibinfo {volume} {20}},\ \bibinfo {pages} {R301}
		(\bibinfo {year} {2003})}\BibitemShut {NoStop}%
\end{thebibliography}
\end{document}